\documentclass{ws-rv975x65}

\usepackage{amsbsy}
\usepackage{amsmath}
\usepackage{amssymb}
\usepackage[english]{babel}

\usepackage{ws-rv-van}

\usepackage{wrapfig}
\usepackage{sidecap}

\usepackage{multirow}

\usepackage[normalem]{ulem}

\usepackage[dvipsnames,usenames]{color}
\definecolor{Red}{rgb}{0.9,0.1,0.1}
\definecolor{Blue}{rgb}{0.0,0.0,0.8}
\definecolor{Green}{rgb}{0.0,0.8,0.0}
\definecolor{Black}{rgb}{0.0,0.0,0.0}

\usepackage{chemarr}

\usepackage{tikz}

\let\captionsize\footnotesize

\renewcommand{\footnotesize}{\renewcommand{\baselinestretch}{1.1}\captionsize\renewcommand{\baselinestretch}{1.0}}

\newcommand{\myemph}{\bgroup\markoverwith{\hbox{\kern-.03em\vtop{\begingroup\kern.2ex\color{Red}\hrule width.2em\kern1.1pt\color{Red}\hrule\endgroup\kern-.03em}}}\ULon}

\renewcommand{\theenumi}{\textbf{\arabic{enumi}}}

\begin{document}


\chapter{More is different: 50 years of nuclear BCS}
\author{R.A. Broglia}
\address{Department of Physics, University of Milano, via Celoria 16, 20133 Milan, Italy}
\address{INFN, Milan Section, Milan, Italy}
\address{The Niels Bohr Institute, University of Copenhagen, Blegdamsvej 17, DK-2100 Copenhagen, Denmark}
\address{FoldLESs S.r.l., via Valosa di sopra 9, I-20052, Monza MB, Italy}


\begin{abstract}

\centerline{Prologue}

At the basis of BCS theory , and associated symmetry breaking phenomena in gauge space, one finds Cooper pair binding. A major 
question in the nuclear case concerning this issue, regards the relative role played by the bare nucleon-nucleon force and by the interaction induced by the exchange of vibrations between members of Cooper pairs. 
The exotic nucleus $^{11}_3$Li$_{8}$ in which two neutrons forming an extended halo, bind  weakly to the $^9$Li core, provides an excellent  testing ground
to try to shed light on this issue. Theory finds that, in this case, the exchange of collective vibrations associated with the core and with  the halo fields, provides an important  fraction of the glue  binding the pair. Inverse kinematics and active detector based experiments, combined with a quantitative  description (based on absolute differential cross sections) of single Cooper pair tunneling, the specific probe of pairing in nuclei, which forces the virtual phonon into a real final state, have tested these predictions with positive results. The extension of structure and reaction studies to open shell (superfluid) nuclei (Sn-isotopes), displaying a strong alignment 
of quasispin in gauge space, and associated domain wall, as testified by pairing rotational bands excited 
in terms of single Cooper pair tunneling, provides an overall description of 
the data within experimental errors. This is also true  in connection with pairing vibrations as observed in closed shell nuclei. 

Many of the concepts which are at the basis of the development associated with a quantitative treatment of the variety of phenomena  associated with the spontaneous breaking of gauge symmetry 
in nuclei have been instrumental in connection with novel studies of soft matter, namely of protein evolution and protein folding. Although the
route to these subjects and associated 
development does not necessarily imply the nuclear physics connection, such a connection has proven qualitatively and quantitatively inspiring.

In particular to model protein evolution  in terms of the alignment  of quasispins displaying twenty different projections, one for each of the twenty amino acids occurring in 
nature, and the associated symmetry breaking  in information (sequence) space. Emergent properties of the corresponding phase transition are domain
walls which stabilize local elementary structures (LES), few groups of 10-20 aminoacids which become  structured already in the denatured state  with varied degree of stability, 
and which can be viewed as (mutually complementary) virtual  secondary structures providing  the molecular recognition directing  protein folding under biological conditions  (solvent, temperature, pH, etc.). In fact, their docking  is closely related to the transition state of the process. 
While the two-step, yes or no, folding process, does not provide direct   information concerning LES,
one can force LES from virtual  to become  real, observable  final state  entities. 
Getting again inspiration from the nuclear case (virtual processes contributing to pair correlations  can be forced to become real with the help of a probe which itself changes  particle number by two), one would expect that to make real virtual LES, that is segments of the protein which already at an early stage of the 
folding process flicker in and out of  their native conformation, one needs a probe which itself displays  a similar behaviour. It is almost self-evident that 
peptides (p-LES$_i$) displaying  a sequence identical to LES$_i$ will bind with high specificity  to its complementary LES$_j$, blocking  folding and exposing the (p-LES)$_i$-LES$_j$
complex to direct observation.
Results  \textit{in vitro} and \textit{ex vivo} (infected cells) of the folding inhibition ability of p-LES, testify to the soundness of the picture. Based again on the analogy with the nuclear case (become theoretically and experimentally quantitative in the prediction and measurements of Cooper-pair 
tunneling), much effort has been concentrated in defining the new  paradigm of protein inhibition, namely folding inhibition, and to develop the in silica protocols, and the
wet laboratory activity assays, which quantitatively can map out  the role LES play in folding.
It is expected  that eventually, this insight  can help in developing  p-LES into leads of non-conventional drugs.

\end{abstract}



\section{Overview}

``More is different'' is the title of a seminal article which Phil Anderson wrote (\cite{Anderson:72} , see also  \cite{Anderson:95}) to oppose the reductionist point of view which reads something like: ``$\ldots$ only scientists who are studying anything really fundamental are those who are working on (fundamental) laws $\ldots$'', to which the author replied: ``$\ldots$ I would challenge you to start from the fundamental laws of quantum mechanics and predict the ammonia molecule inversion (``superconductivity'' he writes in another version of the same subject) $\ldots$'' and he goes then to comment developments concerning nuclear physics (Bohr Mottelson model \cite{Bohr:53}): ``$\ldots$ It is fascinating that $\ldots$ nuclear physicists stopped thinking of the nucleus as a featureless, symmetrical little ball and realized that $\ldots$ it can become football--shaped or plated--shaped. This has observable consequences in the \textbf{reactions} and excitation spectra (\textbf{structure} we would say) that are studied in nuclear physics, even though it is much more difficult to demonstrate directly than the ammonia inversion $\ldots$. Three or four or ten particles whirling about each other do not define a rotating ``plate'' or ``football'' $\ldots$ (but) $\ldots$ \textbf{when we see such spectrum}, even not so separated, and somewhat imperfect, we recognize that the nucleus is, after all, $\ldots$, approaching macroscopic behavior'', (bold face are of RAB).

These words embody one of the best descriptions of the quest to understand the physics which is at the basis of Finite Many--Body systems (FMBS) at large, and of atomic nuclei in particular.
Within this context let us quote from the lectures Ben Mottelson held at Varenna on Lake Como, in 1960 \cite{Mottelson_Varenna}: 
"in ... a many-body system such as a nucleus, every feature  is in some sense a collective phenomenon ... Indeed the most striking and fundamental collective feature in all
nuclear phenomena is  the existence of an average nuclear field in which  nucleons move approximately independently. It is a rather unfortunate perversity of the popular terminology
that regards the collective field as in some sense an antithesis to the nuclear collective effects". For then continuing: "... The typical situation in the middle of a major shell is one in which there are  many near-lying configurations, and because of the (2j+1) degeneracy of each single-particle orbit in any configuration the effective degeneracy in any many-particle  configuration becomes absolutely staggering ... we can say that these many different states correspond to the many  ways of correlating the particles with each other ... The central problem of nuclear spectroscopy might be in terms of the question, which of  these many states, or which linear combination of them, correspond to the most efficient 
correlation of the nucleons ....".

Two schemes  are then discussed: the aligned and the pairing scheme. Or using  the quasispin language, the aligned scheme in 3D- and in gauge space.                                                
Within this scenario,  the (Mayer-Jensen)-(Nilsson) and particle-quasiparticle transformations diagonalize the corresponding mean fields. These mean fields violate rotational and gauge invariance respectively. Taking into account 
the (weak) residual interaction acting between the particle (quasiparticle) states,  the dynamical aspects of the nucleus  can also be described  within the 
framework of the above scheme. As explained by  Aage Bohr in his contribution to the Nuclear International Congress of Paris  of 1964 \cite{Bohr_Paris}:
" ... it has been possible to interpret a very extensive body of evidence on the nuclear spectra in terms  of quite a simple picture, involving just a few types of excitations, the elementary  modes of nuclear excitation. These may be grouped into three classes: particles (or quasi-particles), vibrations, and rotations". 
In other words, the elementary modes of nuclear excitation incorporate  a large fraction of the nuclear correlation, the corresponding product  basis diagonalizing 
to a large extent  the nuclear Hamiltonian.  

Within this context, the LES-conjecture is able to relate  and interpret  a large variety of phenomena  which are at the basis of protein  folding and stability. 
Furthermore, representing the force field  (interaction among residues)  instead than in terms of the individual amino acides, in terms of native structures (NS), i.e. segments of 10-15 residues in their native conformation, of which few (2-3)
are LES,  one goes from   a $N_a \times N_a$ dense matrix ($N_a$: number of amino acids forming the protein, namely a number  of the order of hundreds) to a $n \times n$ co-diagonal matrix (where $n$ is less than ten), provided the different NS, in particular   those corresponding  to the LES, are set  in appropriate  sequence, i.e. following  not the linear, but the 
3D-native like order. This is in particular  required  by the fact that  the native docking  of LES leads to the  post critical Folding Nucleus $(FN)_{pc}$, the minimum set  of native contacts  which inevitably grows into the native  conformation. Testifying to the validity  of this "solution" of the protein folding problem is the fact that  it provides the basic elements of a protocol, with a straightforward  software embodiment, and with an eventual realization based on Protein (sequence) Data Base (PDB), which allows to individuate the protein segments associated with LES, and thus the possible candidates to p-LES. 

This is important to shed light  on the protein folding process,  but also in connection with  technological transfer  (spin off) perspectives. In fact, intervening in the folding 
with p-LES can lead to a nonproductive folding, and thus to non expressed pathogen agents (this phenomenon is in a way similar to the anti-pairing  effect observed in
atomic nuclei when blocking, with an odd nucleon, a hot  orbital contributing to the nuclear pairing gap). Because the amino acids which stabilize the LES play an essential role in the folding process, p-LES can be viewed  as potential leads to drugs which do not elicit resistance, arguably defining a new paradigm to fight disease at large, and infective disease in particular. 

Much effort is being invested in developing the proof of concept of non-conventional inhibition \cite{inhibition}. In particular, improving  the value of the dissociation constant $K_D$ which measures the strength with which a p-LES binds its complementary segment in the target protein, a quantity  which plays, within the protein folding context, a similar  role $S_{2n}$ does in the (halo Cooper pair)-($^9$Li core) binding scenario. Because proteins are, as a rule, folded proteins, exception made when they are expressed  and before becoming mature, 
the relation of $K_D$ and the inhibition constant $K_I$, which is  straightforward in the case of active site centered  conventional inhibitors (molecules which resemble the substrate binding to the active site), is less than obvious in the case of folding inhibitors. This is in keeping with the fact that little is known  concerning the probability  that 
a complementary LES segment to a p-LES is exposed to  the solvent in the different situations (cells, in vitro, etc.). Consequently, while conventional inhibition controls are available (like e.g. ATV in the case of the HIV-1-protease), no drug which blocks folding  exists, let alone a reference molecule. In other words, regarding
research aimed at studying and measuring how symmetry in information space is concretely broken in a protein, and its eventual application to design folding inhibitors,  one is in a situation which resembles that  in which scientists found themselves in trying to learn about phase coherence in gauge space, before the discovery of the  Josephson effect.

Summing up,  concepts developed in nuclear studies of spontaneous symmetry breaking of gauge invariance and emergent properties (generalized rigidity in gauge space) in a regime of strong fluctuations (pairing vibrations), many of which stem from Phil Anderson's own work, have found their way in helping to explain some aspects of the becoming as well as of the dynamical behaviour (folding) of other FMBS like e.g. proteins. Arguably, one can posit that: 
\emph{not only more is different, but more is also simpler, and in a very real sense, more is more fundamental (than the basic laws)}. Emergent properties, that is properties not contained in the particles forming the system under study, nor in the interactions acting among them, will essentially decide many fundamental issues, eventually  also in a universe with different particles and interactions than the ones we know\footnote{Within this context, think of how little relevant was to know all about the psyche of each individual to predict their behaviour when assisting to a rock concert or a soccer match (if we replace individual by Mitteleuropeans and concert or game with ``political'' manifestations in Munich and Berlin, we have Elias Canetti's Mass und Macht (Mass and Power)), testifying to the fact that individual relations under such circumstances were so strongly renormalized by the context that new, completely unexpected behavioural patterns substituted the well known, in average highly civilized patterns known in the daily interweaving of the lives of single individuals.}.

As we shall have the chance to comment at the end of the present narrative, conspicuous evidence for the validity of the above statement stems from the (partial) answer to a question Phil Anderson puts forward at the end of his paper written together with L. Stein \cite{Anderson:84} ``\emph{Broken symmetry, emergent properties, dissipative structures, life: are they related ?}''
The concepts of elementary modes of excitation introduced by Landau, together with the flexible rules for picturing what cannot be pictured, namely virtual states, known as Feynman diagrams\footnote{It happened $\ldots$ in the summer of 1948 (car travel from Cornell to Los Alamos)\\
``$\ldots$ Dick distrusted my mathematics and I distrusted his intuition $\ldots$''\\
``$\ldots$ Dick was right $\ldots$ because his sum--over--histories theory (Feynman path integrals) $\ldots$ was solidly rooted in physical reality $\ldots$''\\
As a consequence ``$\ldots$ Dick's flexible rules, now known as Feynman diagrams, are the first working tool of every theorist''\\
\vspace{0.5cm}
F. Dyson, Disturbing the Universe, Harper, New York (1979)} are at the basis of Nuclear Field Theory (NFT)\cite{Bes:76,Bes:76b,Bes:76c,Broglia:76,Bes:75,Bortignon:77}. It provides the rules to work out the interweaving of collective and single--particle degrees of freedom in atomic nuclei. In much the same way as in QED \cite{Tomonaga:46,Schwinger:48,Schwinger:48b,Feynman:49,Feynman:49b,Feynman:50,Dyson:49} , the corresponding couplings and associated renormalization effects arize, in the nuclear case from virtual processes, associated with vacuum (ground state) fluctuations, processes which can become real in the presence of an external field (probe).
A typical example of these phenomena in QED is provided by the Lamb shift, namely the splitting between the s,p--orbitals in the hydrogen atom, a phenomenon closely related with the Pauli principle (see Fig. 1).

\begin{figure}[h]
	\begin{center}
		\includegraphics[width=0.6\textwidth]{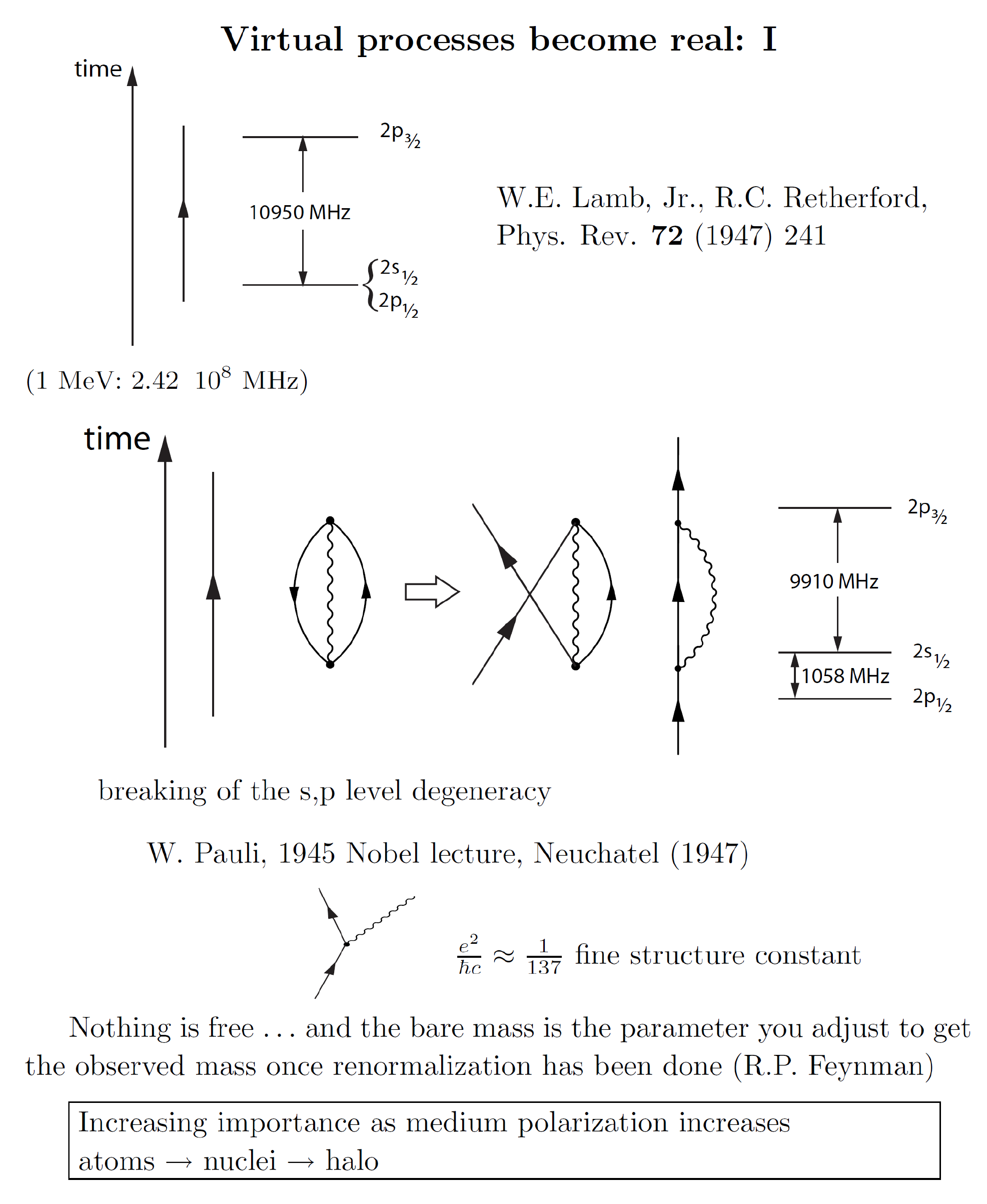}
		\caption{Schematic resum\'{e} of the processes which are at the basis of the Lamb shift, in terms of Feynman diagrams involving the coupling between photons, electrons and positrons.}
	\end{center}
\end{figure}

\section{A single Cooper pair nuclear system: ${}^{11}$Li and the single pair tunneling reaction ${}^{11}$Li$(p,t)$ ${}^9$Li}

The role of photons in QED is played, in nuclei, by collective modes. In the case of the single Cooper pair system ${}^{11}$Li, especially by the pigmy resonance, namely a low--lying isovector dipole vibration. This is a chunk of the GDR of the core ${}^9$Li in which protons and neutrons move out of phase, a mode which is intimately related to the spontaneous symmetry breaking of space homogeneity associated with the fact that the center of mass of a finite system like the atomic nucleus, specifies a privileged position in space (see Fig. 2). 


\begin{figure}[h!]
	\begin{center}
		\includegraphics[width=0.7\textwidth]{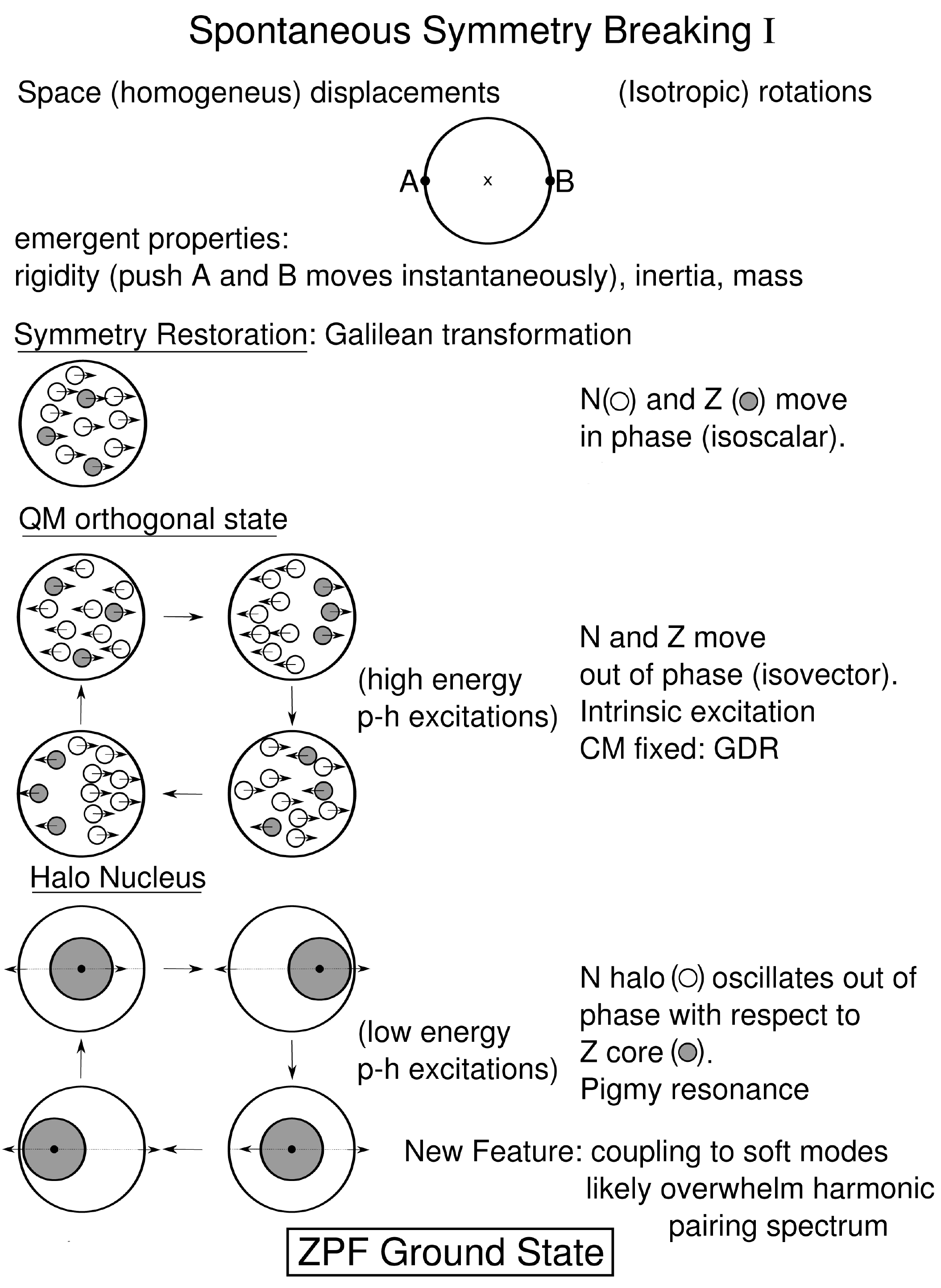}
		\caption{Schematic representation of the breaking of translational invariance, pushing model restoration (see ref. \cite{Bohr:75}), and associated orthogonal states (dipole GR and pigmy resonance).}
	\end{center}
\end{figure}

While ${}^9_3$Li${}_6$ is bound, ${}^{10}_3$Li${}_7$ is not. Nonetheless the single--particle $s_{1/2}$ and $p_{1/2}$ resonances of this system have been studied in detail. It is found that through renormalization processes connected to Pauli-principle-like diagrams as those encountered in connection with the atomic Lamb shift, the $p_{1/2}$ is shifted to higher energy from that predicted by a standard mean field potential, in keeping with the fact that the quadrupole vibration of the ${}^8$He core has as main component the $(p_{1/2},p_{3/2}^{-1})_{2^{+}}$ configuration. Coupling of the $s_{1/2}$ continuum state to the different vibrations of the core lowers the energy of this state (resonance) to energies close, but below that of the $p_{1/2}$ state (parity inversion leading, among other things, to a new ``magic number'', namely $N=6$ shell closure, see Fig. 3(I)).

Within this scenario ${}^{11}$Li(gs) corresponds to an unbound $s_{1/2}^2 (0)$ configuration (see Fig. 3(II)). The bare residual interaction lowers this configuration by less than 100 keV. On the other hand the exchange of the quadrupole mode of the ${}^9$Li core and of the pigmy resonance of ${}^{11}$Li lead to a neutron Cooper pair bound by about 300 keV, the experimental value being $\approx 380$ keV. This neutron halo state is the pair addition mode of the N=6, $^{9}$Li closed shell system. \textit{Of notice that the pigmy resonance is the result of a delicate (Baron M\"unchausen--like) bootstrap process, in which an originally extended neutron halo created by the two unbound neutrons passing by ${}^9$Li are, quantum mechanically, forced to slosh back and forth with respect to the proton core field (ZPF), leading to a collective mode which ,exchanged between the halo neutrons, binds the Cooper pair to the core}. In other words, the pigmy resonance is in a very real sense a consequence of (translational) symmetry restoration and of a virtual process (vibrations of an extended neutron field) becoming real as a low--lying excitation, after having acted as glue to bind the two outer neutrons to the ${}^{9}$Li core thus generating the weakly, but nonetheless bound ground state of ${}^{11}$Li (see Fig. 2, lower part, and Fig. 3(II); see also ref.\cite{epj_li11,Brink:10} and refs. therein).

We are then in presence of a paradigmatic nuclear embodiment of Cooper's model\cite{Cooper:56} which is at the basis of BCS theory: a single weakly bound neutron pair on top of the Fermi surface of the ${}^9$Li core. But the analogy goes beyond these aspects, and covers also the very nature of the interaction acting between Cooper pair partners. Because of the high polarizability of the system under study, most of the Cooper pair correlation energy stems, according to NFT (see \cite{epj_li11} and refs. in the caption to Fig. 3), from the exchange of collective vibrations, the role of the bare interaction being, in this case, minor. \textit{In other words, we are in the presence of a new realization of Cooper's model in which a totally novel Bardeen--Pines--like phonon induced interaction\cite{Bardeen:55}, is generated by a self induced collective vibration of the nuclear medium}. Because one is in possess of the specific tool to probe pairing correlations in nuclei, namely, two--particle transfer reactions (see contribution Potel and Broglia to this Volume), one can force these virtual processes to become real (see Fig. 4).


Making use of the extension of NFT to reactions \cite{Broglia:05} (continuum states, already present in embryo in the calculation of the single--particle resonant states of ${}^{10}$Li (see Fig. 3 (I))), one can calculate the absolute ${}^{11}$Li$(p,t) \;{}^9$Li($p_{3/2}(\pi) \otimes 2^+ ({}^8$He$); 1/2^{-})$ cross section to the first excited state of ${}^9$Li, that is, to the lowest member of the multiplet of states arizing from the coupling of the ${}^8$He quadrupole vibration to the $p_{3/2}$ proton state. To do this one has to take properly into account the successive and simultaneous contributions to the transfer amplitude, corrected because of  non--orthogonality effects. As a rule, and exception made for 
Q--value effects in which single--particle channels become closed
 (see e.g.\cite{Potel:11,Wimmer:10} for recent references), the successive contribution to the two--particle transfer cross section is the dominant one, non--orthogonality canceling much of the already weak, simultaneous contribution. Of notice that similar issues were debated in connection with the proposal of Josephson \cite{Josephson:62} concerning the possibility of observing a supercurrent across a dioxide layer separating two superconductors, and Bardeen's objection that the pairing gap is zero inside the layer \cite{Bardeen:62}. The answer to such an objection is to be found in the fact that it is $\alpha_0$ ($=\langle P ^{\dagger} \rangle $) which controls tunneling and not $\Delta$, a fact that emerges naturally from Gorkov's formulation of superconductivity (see contribution of Potel and Broglia to the present Volume).

Within the framework of nuclear reactions, one is dealing, as a rule, with normal--normal and normal--superfluid tunneling. In particular this last situation within the framework of condensed matter was taken up by Cohen, Falicov and Phillips in connection with the Josephson--Bardeen discussion (\cite{Josephson:62,Bardeen:62,Cohen:62} , see also \cite{Anderson:63} ).



\begin{center}[h]
	\hspace{-1cm}\includegraphics[width=0.7\textwidth]{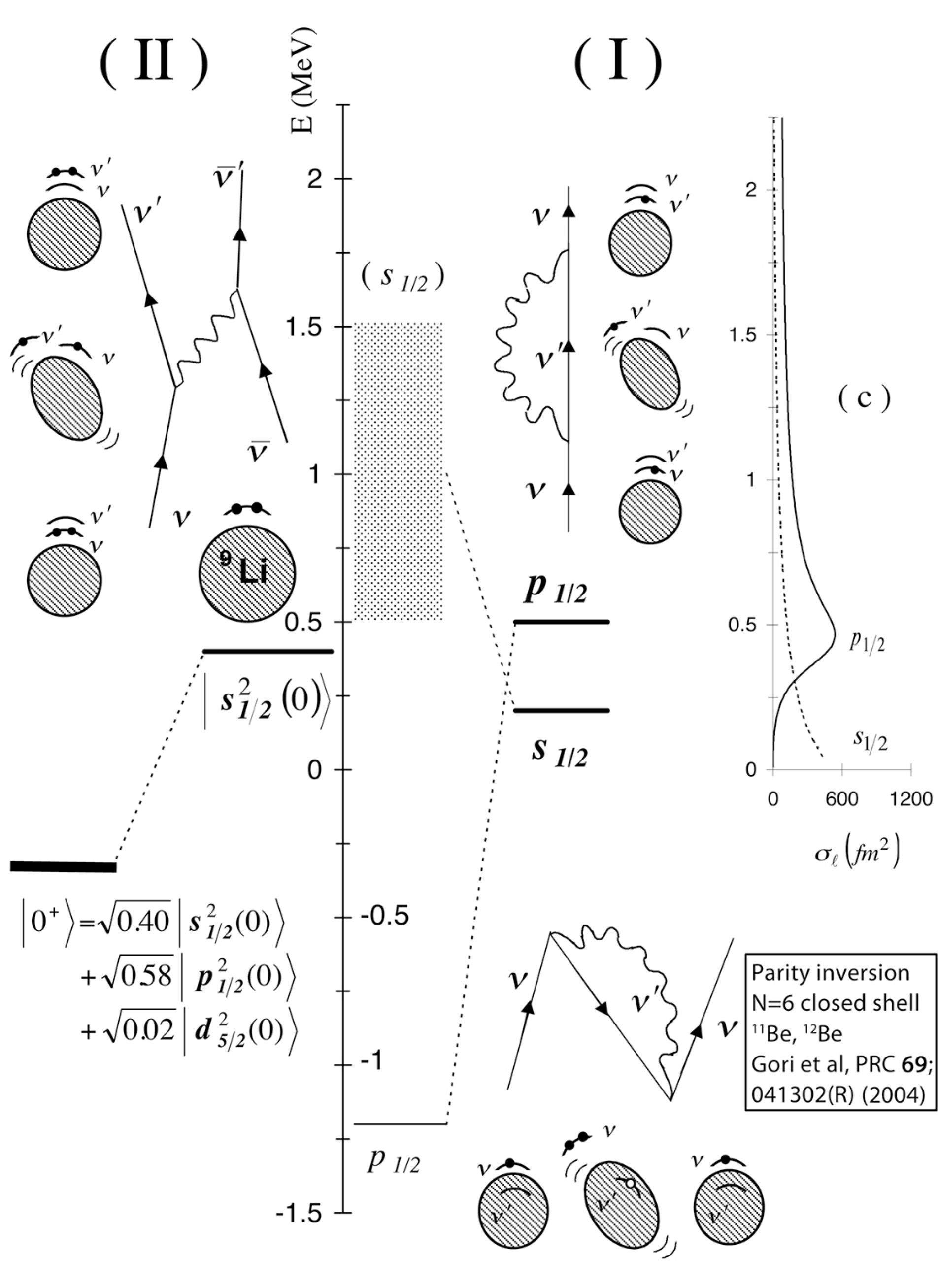}
\end{center}
\vspace{-0.5cm}
\begin{center}
 		Barranco et al., EPJ \textbf{A11} (2001) 305
\end{center}
\begin{footnotesize}
	\begin{center}
		Figure 3. Schematic representation of the dressing of single-particle and induced pairing interaction in $^{10}$Li (I), and $^{11}$Li (II), (reported with permission from Barranco et al. Eur. Phys. J. \textbf{A11} (2001) 305, Copyright 2001, European Physical Journal)
	\end{center}
\end{footnotesize}

The NFT description of the single Cooper pair system ${}^{11}$Li summarized in Fig. 3 together with the NFT reaction description of ${}^{11}$Li$(p,t) \;{}^9$Li reaction (Fig. 4), provides
\cite{Potel:10} an accurate account of the experimental findings\cite{Tanihata:08} . In particular, direct evidence for phonon mediated pairing in nuclei (see Fig. 5 of the contribution of Potel and Broglia to this Volume). At variance with the case of the infinite system (e.g. normal superconducting metals) in which there is a bound state for any strength of the interaction, in finite FMBS there is a lower limit for the strength below which the system correlates but does not condense. This is what happens around closed shell nuclei, in which the decoupling between occupied and empty states blocking pair condensation, arizes from the gap in the single--particle spectrum observed at magic numbers, and forced upon the system by the ``external'' mean field produced by all the nucleons on the motion of each single neutron and proton. As a consequence, pairing vibrations (see upper part Fig. 5) are, in atomic nuclei, strongly excited in single Cooper pair transfer reactions.

\begin{center}\begin{large}\textbf{Virtual processes become real: II}\end{large}\end{center}

\vspace{0.3cm}

\includegraphics[width=0.534\textwidth]{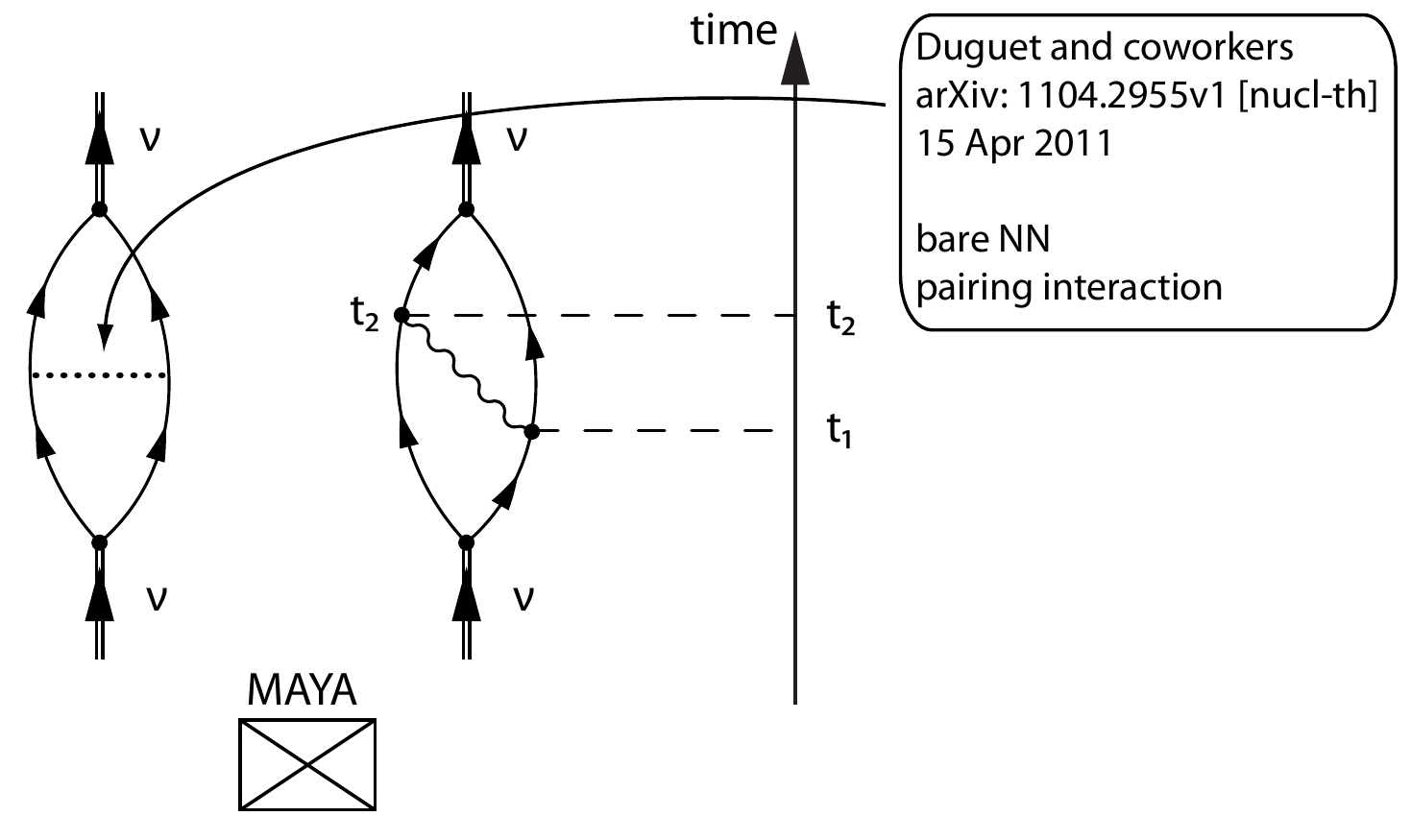} \hspace{0.1cm} \raisebox{2cm}{\begin{minipage}[c]{0.25\textwidth} \begin{small} F. Barranco et al., \\ Eur. Phys. J. A \textbf{11}, 385 (2001) \end{small} \end{minipage}}

\vspace{0.3cm}

\includegraphics[width=0.558\textwidth]{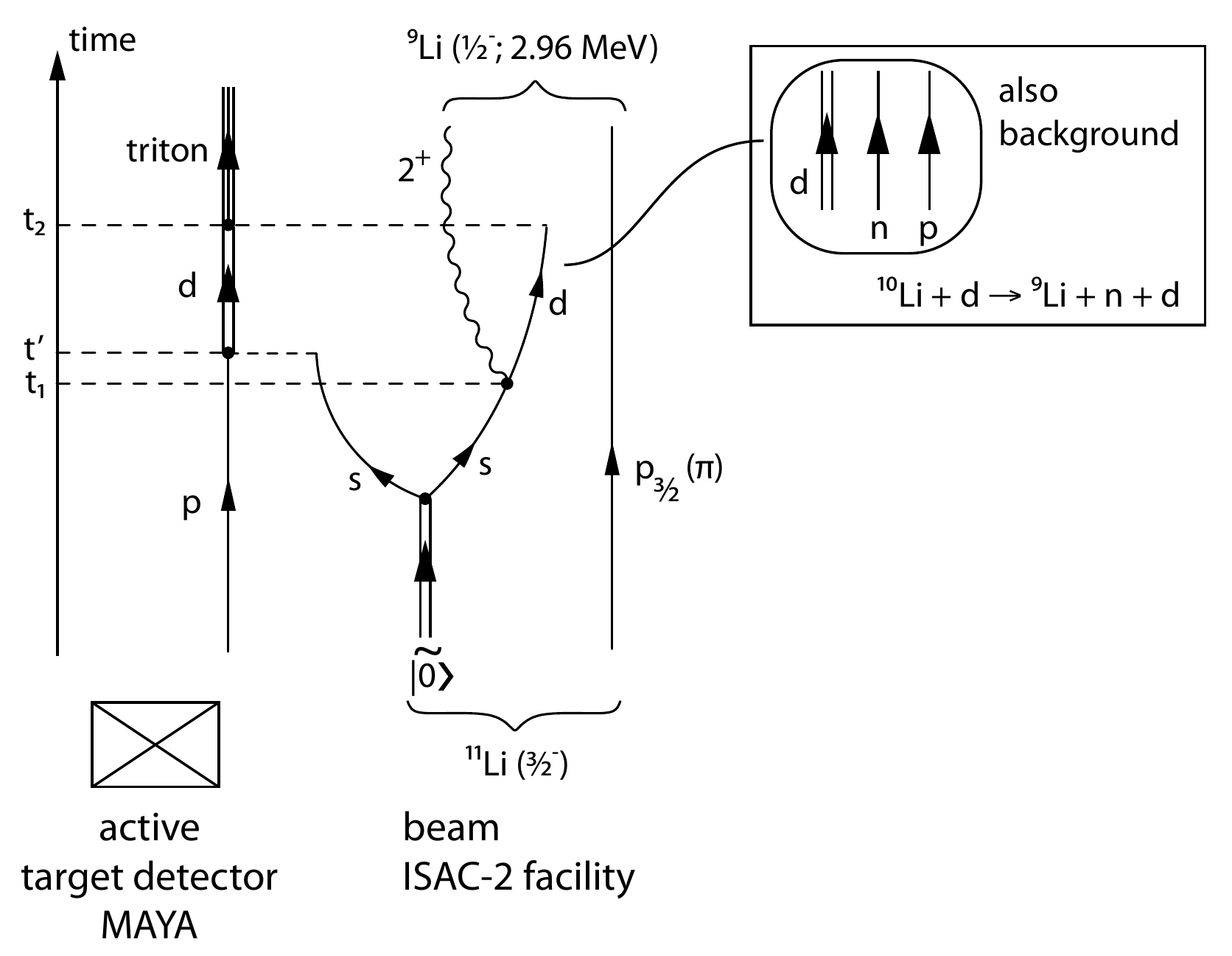} \hspace{0.1cm}\raisebox{2cm}{\begin{minipage}[c]{0.37\textwidth} \begin{small} G. Potel et al., \\ PRL \textbf{105}, 172502 (2010) \vspace{0.5cm} \\ D.M. Brink and R.A. Broglia \\ Nuclear Superfluidity, \\ Cambridge University Press, \\ Cambridge (2005) \vspace{0.5cm} \\ I. Tanihata et al \\ PRL \textbf{100}, 192502 (2008) \end{small} \end{minipage}}

\vspace{-0.2cm}
\begin{footnotesize}
	\begin{center}
		Figure 4. Schematic representation of the bare nucleon-nucleon and phonon induced pairing correlations (upper part) NFT diagrams, and of the excitation of the final, excited state of $^{9}$Li($1/2^{-}; 2.69$ MeV), in the TRIUMF experiment  reported in ref. \cite{Tanihata:08} (see also \cite{Potel:10}).
	\end{center}
\end{footnotesize}

\setcounter{figure}{4}
\begin{figure}[h!]
	\begin{center}
		\includegraphics[width=0.5\textwidth]{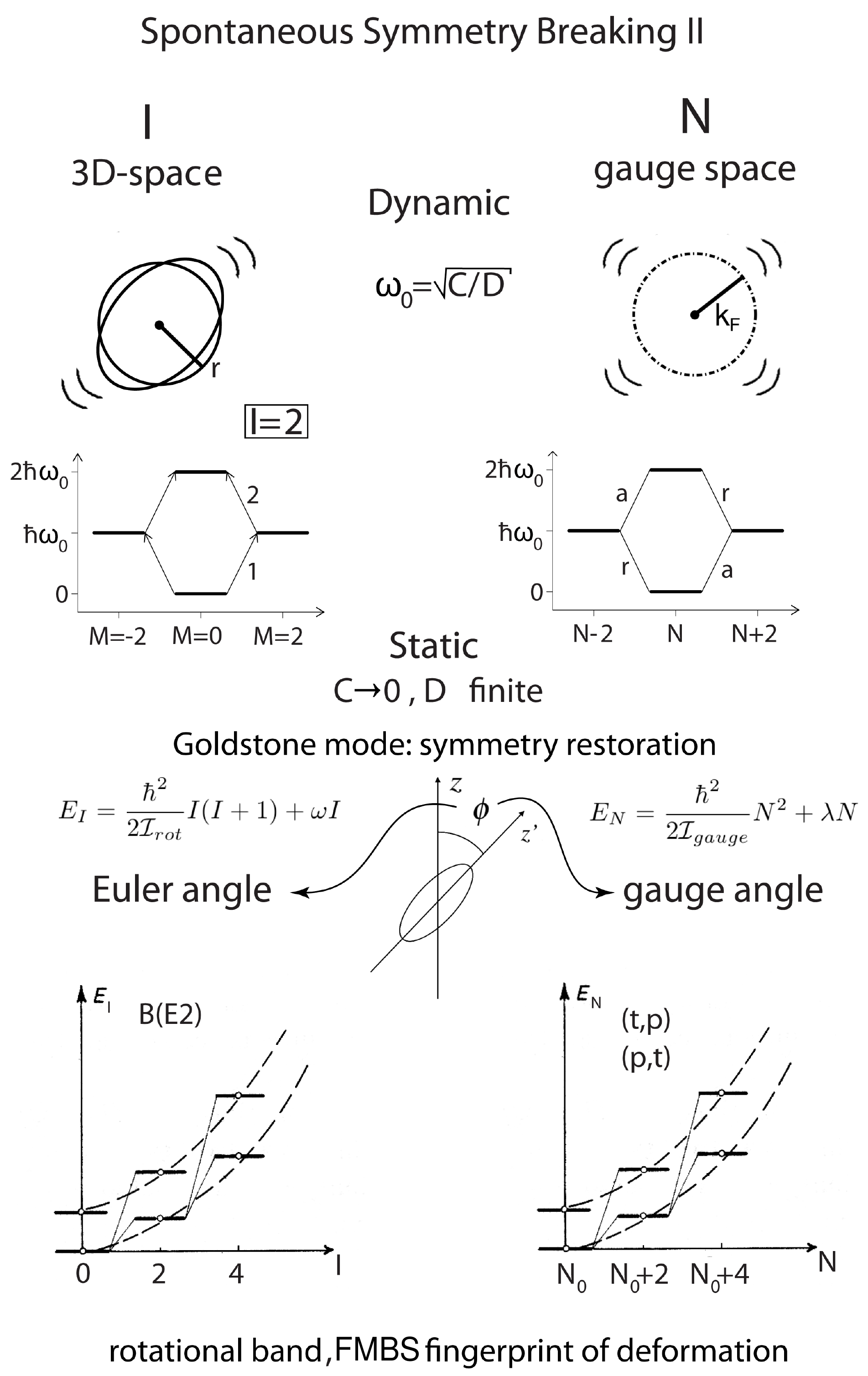}
	\caption{Schematic representation of collective modes associated with dynamical and static distortions violating rotational and gauge symmetries (see also table XI in ref. \cite{Broglia:73})}
	\end{center}
\end{figure}

Now, away from closed shells (open shell nuclei), such a (large) single--particle gap disappears, and one is left with rather modest differences in energy between occupied and empty states. Under such circumstances, Cooper pairs condense, the system becomes superfluid, and BCS theory provides a good description of nuclear properties. In particular the fact that the mixing between occupied and empty states gives rise to a privileged orientation in gauge space, and thus to particle number violation. The observation of pairing rotational bands (see lower part Fig. 5) being the fingerprint of nuclear spontaneous symmetry breaking in gauge space (see the Chapters of this Volume contributed by Bes, Hansen and Potel and Broglia).

\section{Pairing vibrations in light exotic nuclei}

Because of the parity inversion discussed in Sect. 2 (see also Fig. 3 (I)) in connection with ${}^{10}$Li and ${}^{11}$Li, $N=6$ becomes, for this neutron drip nucleus, a magic number instead of $N=8$. Consequently, one can expect a pairing vibrational spectrum for ${}^9_3$Li$_6$, but also for ${}^{10}_4$Be$_6$ as shown in Fig. 6, where the harmonic predictions are compared with the available experimental findings. Major anharmonicities are expected in connection with the 5.7 MeV $|gs({}^{11}$Li$)\otimes gs ({}^7$Li$)\rangle$, two--phonon, pairing vibrational state of ${}^9$Li, in keeping with the high polarizability of ${}^{11}$Li.

Concerning the pairing vibrational spectrum of ${}^{10}$Be, experiments which could test some of the predictions collected in Fig. 6, are planned to be carried out \cite{Kanungo:prop} at TRIUMF (see also the contribution of Kanungo and Tanihata to this Volume).
\setcounter{figure}{5}
\begin{figure}[hbt!]
	\begin{center}
		\includegraphics[width=0.44\textwidth]{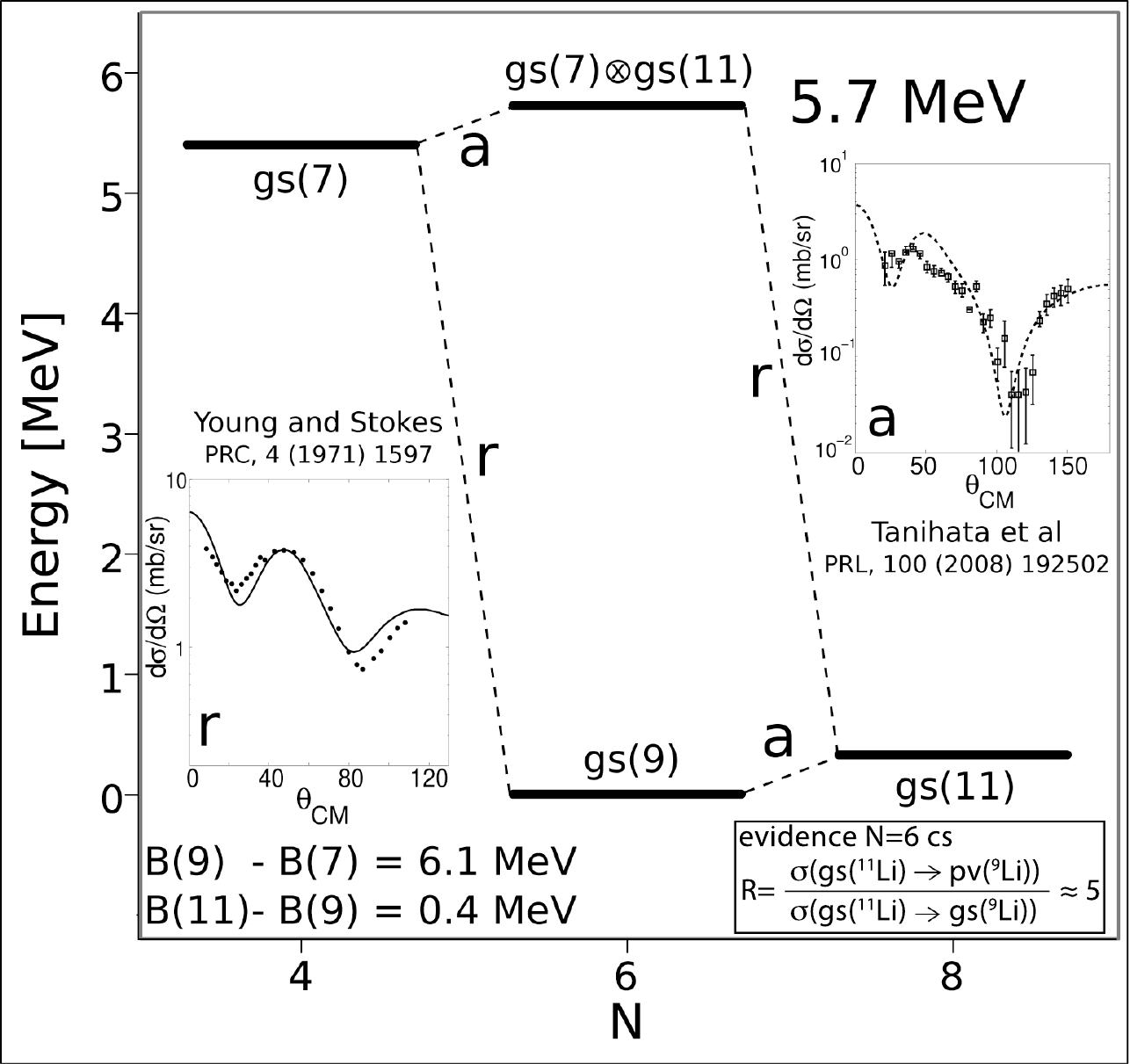} \quad		\includegraphics[width=0.44\textwidth]{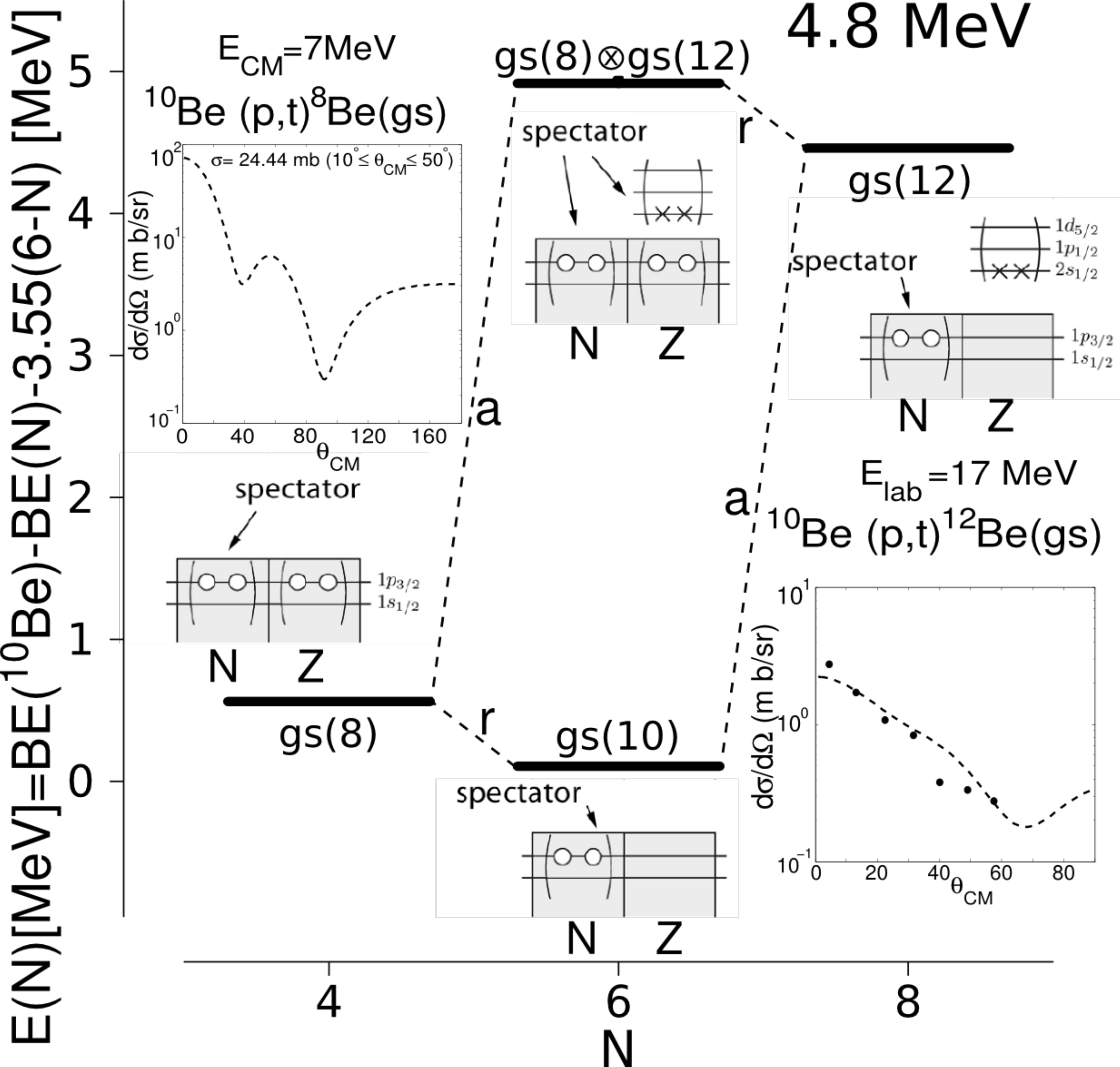}

		\caption{Pairing vibrational spectrum of $^9$Li (Cf. ref. \cite{Potel:arxiv2}) (left). Pairing vibrational spectrum of $^{10}$Be (Cf. ref. \cite{Potel:arxiv2}) (right). The data is from \cite{Tanihata:08} and from Fortune et al. (see Caption to Table 1) respectively.}
	\end{center}
\end{figure}




\section{Many--Cooper pair systems: Sn--isotopes}

The sequence of the Sn--isotopes extends from closed shell to closed shell systems, namely, from ${}^{100}_{50}$Sn$_5$ to ${}^{132}_{50}$Sn$_{82}$ and beyond. It can, in principle, provide important information concerning the transition between the pairing vibrational regime typical of normal systems, to the pairing rotational regime, representative of condensed, superfluid nuclei.

While a well developed pairing rotational band is formed by the ground states of the Sn--isotopes, consistent deviations from such a picture are observed, as expected, around the end points, namely around closed shell systems. 

\section{Level of accuracy of pairing vibrational and rotational studies with two--nucleon transfer reactions}

Absolute two--particle transfer cross sections are to be calculated for a quantitatively accurate comparison between theoretical and experimental studies of nuclear pairing. This is possible by treating on equal footing, for example within the framework of NFT, both structure and reactions (see \cite{Bes:76,Bes:76b,Bes:76c,Broglia:76,Bes:75,Bortignon:77} and \cite{Potel:11,Potel:10,Broglia:05,Potel:arxiv2};
see also the contribution of Thompson to the present Volume), and making use of global sets of optical parameters. No details are given in the present contribution
as to  the methods, routines and theoretical details entering the calculations, as they have been extensively reported in the literature quoted above (see also the contribution of Potel and Broglia to this Volume). 
Examples of such a program are collected in Table 1. From these results, as well as many detailed controls\cite{Potel:arxiv2} , it seems fair to summarize the present situation concerning the probing of single- and multi-Cooper pair nuclei with the help of two-nucleon transfer reactions
as written in the conclusions. 


\begin{table}[h!]
\tbl{Calculated \cite{Potel:11,Potel:10,Potel:arxiv2} absolute two-nucleon transfer cross section in comparison with the experimental findings (see $^{e)-k)}$).}
{  \begin{tabular}{|c|c|c|c|}
\cline{2-4} 
\multicolumn{1}{c|}{}                                                                & \multicolumn{3}{|c|}{$\sigma($gs$\rightarrow$f)}              \\
\cline{2-4}
\multicolumn{1}{c|}{}                                                                & f        & Theory${}^{d,k)}$         & Experiment${}^{e-j)}$                        \\
\hline 
$^{7}$Li($t,p$)$^{9}$Li                                                              & gs       & 2.3 ${}^{a)}$  & $1.9 \pm 0.57$ ${}^{a)}$              \\
\hline
\multirow{2}{*}{${}^{11}$Li$\left({}^{1}\textrm{H},{}^{3}\textrm{H}\right){}^{9}$Li} & gs       & 6.1 ${}^{a)}$  & $5.7 \pm 0.9$ ${}^{a)}$           \\
\cline{2-4}
                                                                                     & $1/2^{-}$& 0.7 ${}^{a)}$  & $1.0 \pm 0.36$ ${}^{a)}$          \\
\hline
$^{10}$Be($t,p$)$^{12}$Be                                                            & gs       & 14.3 ${}^{a)}$  & $14.7 \pm 4.4$ ${}^{a)}$            \\

\hline
\hline 
$^{112}$Sn($p,t$)$^{110}$Sn, $E_{CM}=26$ MeV                                            & gs       & 1301 ${}^{b)}$           & $1309 \pm 200 (\pm 14)$ ${}^{b)}$\\
\hline 
$^{116}$Sn($p,t$)$^{114}$Sn, $E_{CM}=26$ MeV                                            & gs       & 2078 ${}^{b)}$           & $2492 \pm 374 (\pm 32)$ ${}^{b)}$\\
\hline 
$^{118}$Sn($p,t$)$^{116}$Sn, $E_{CM}=24.4$ MeV                                            & gs       & 1304 ${}^{b)}$           & $1345 \pm 202 (\pm 24)$ ${}^{b)}$\\
\hline 
$^{120}$Sn($p,t$)$^{118}$Sn, $E_{CM}=21$ MeV                                            & gs       & 2190 ${}^{b)}$           & $2250 \pm 338 (\pm 14)$ ${}^{b)}$\\
\hline
${}^{122}\textrm{Sn}(p,t){}^{120}$Sn, $E_{CM}=26$ MeV                                   & gs       & 2466 ${}^{b)}$           & $2505 \pm 376 (\pm 18)$ ${}^{b)}$\\
\hline 
$^{124}$Sn($p,t$)$^{122}$Sn, $E_{CM}=25$ MeV                                            & gs       & 838  ${}^{b)}$           & $958 \pm 144 (\pm 15)$ ${}^{b)}$\\
\hline
\hline 
$^{112}$Sn($p,t$)$^{110}$Sn, $E_p=40$ MeV                                             & gs       & 3349 ${}^{c)}$ & $3715 \pm 1114$ ${}^{c)}$ \\
\hline
$^{114}$Sn($p,t$)$^{112}$Sn, $E_p=40$ MeV                                             & gs       & 3790 ${}^{c)}$ & $3776 \pm 1132$ ${}^{c)}$ \\
\hline
$^{116}$Sn($p,t$)$^{114}$Sn, $E_p=40$ MeV                                             & gs       & 3085 ${}^{c)}$ & $3135 \pm 940$ ${}^{c)}$ \\
\hline
$^{118}$Sn($p,t$)$^{116}$Sn, $E_p=40$ MeV                                             & gs       & 2563 ${}^{c)}$ & $2294 \pm 668$ ${}^{c)}$ \\
\hline
$^{120}$Sn($p,t$)$^{118}$Sn, $E_p=40$ MeV                                             & gs       & 3224 ${}^{c)}$ & $3024 \pm 907$ ${}^{c)}$ \\
\hline
$^{122}$Sn($p,t$)$^{120}$Sn, $E_p=40$ MeV                                             & gs       & 2339 ${}^{c)}$ & $2907 \pm 872$ ${}^{c)}$ \\
\hline
$^{124}$Sn($p,t$)$^{122}$Sn, $E_p=40$ MeV                                             & gs       & 1954 ${}^{c)}$ & $2558 \pm 767$ ${}^{c)}$ \\
\hline 
\hline
$^{206}$Pb($t,p$)$^{208}$Pb                                                          & gs       & 0.52 ${}^{a)}$ & $0.68 \pm 0.21$ ${}^{a)}$         \\
\hline
$^{208}$Pb($^{16}$O,$^{18}$O)$^{206}$Pb                                              & gs       & 0.80 ${}^{a)}$  & $0.76 \pm 0.18$ ${}^{a)}$         \\
\hline
  \end{tabular} }
 \begin{tabnote}
   It is of notice that the number in parenthesis (last column) corresponds to the statistical errors.

   ${}^{a)}$ mb
    
   ${}^{b)}$ $\mu$b

   ${}^{c)}$ $\mu$b/sr ($\sum_{i=1}^{N}(d\sigma/d\Omega)$; differential cross section summed over the few, $N=3-7$ experimental points)

   ${}^{d)}$ Potel et al arXiv: 0906.4298v3 [nucl-th]

   ${}^{e)}$ P. Guazzoni, L. Zetta, et al., Phys. Rev. \textbf{C 60}, 054603 (1999).
 
    P. Guazzoni, L. Zetta, et al., Phys. Rev. \textbf{C 69}, 024619 (2004).
    
    P. Guazzoni, L. Zetta, et al., Phys. Rev. \textbf{C 74}, 054605 (2006).
    
    P. Guazzoni, L. Zetta, et al., Phys. Rev. \textbf{C 83}, 044614 (2011).
    
    P. Guazzoni, L. Zetta, et al., Phys. Rev. \textbf{C 78}, 064608 (2008).

   ${}^{f)}$ G. Bassani et al. Phys. Rev. \textbf{139}, (1965)B830.

   ${}^{g)}$ P.G. Young and R.H. Stokes, Phys. Rev. \textbf{C 4}, (1971) 1597
    
   ${}^{h)}$ H.T. Fortune, G.B. Liu and D.E. Alburger, Phys. Rev. \textbf{C 50}, (1994) 1355

   ${}^{i)}$ J.H. Bjerregaard et al., Nucl. Phys. \textbf{89}, (1964) 337

   ${}^{j)}$ J.H. Bjerregaard et al., Nucl. Phys. \textbf{A 113}, (1968) 484

   ${}^{k)}$ B. Bayman and J. Chen, Phys. Rev. \textbf{C 26} (1982) 1509 and refs. therein
 \end{tabnote}
\end{table}

\section{Conclusions}

The results presented above are likely to signal, if not the starting of the ``exact'' era of nuclear pairing studies, making use of the specific probe provided by single Cooper pair tunneling, in any case the end of the qualitative one which was mainly based on relative two--particle transfer reaction cross section calculations in confronting theory with experiment.

\section{Hindsight}

Let us now come back to the statement made in the overview, that more is not only different but likely more fundamental. Within this scenario one can venture to posit that the actual power of reductionism is to be found not so much in the search of a single particle, but in the simplicity of the emergent properties of complex systems, as a result of a variety of
phenomena of spontaneous symmetry breaking. For example, that associated with information space (amino acid sequence in a polymer) and the origin of proteins\cite{Broglia:arxiv} (and thus metabolism and eventually life on earth). In other words, the fact that all the information needed to code for the three-dimensional structure of a biologically active, native protein (secondary, tertiary and eventually, in the case of multimers, quaternary structures) in its normal physiological milieu (solvent, pH, ionic strength, temperature and other) is coded in the linear, amino acid sequence (primary structure), in a given environment (see \cite{Anfinsen:73} and refs. therein). And this is so, in spite of the fact that such primary structures, apparently, show no regularity, special features, or particular characteristics (cf. \cite{Sanger:52,Fehrst:99} and refs. therein).

Within this scenario let us return to Phil Anderson's paper on the origin of life\cite{Anderson:84}, in particular to his most prominent question, namely: ``\emph{$\ldots$ how does one assign a meaningful information content to a polymer? So far we've only discussed necessary $\ldots$ conditions for symmetry breaking `in information space' to occur. One would guess that, in some sense, \textbf{structure} and \textbf{function} are intimately related $\ldots$ This is one of the most fundamental problems in understanding the origin of life $\ldots$}'' (bold face are of RAB).


\setcounter{figure}{6}
\begin{figure}[h]
	\begin{center}
		\includegraphics[width=0.45\textwidth]{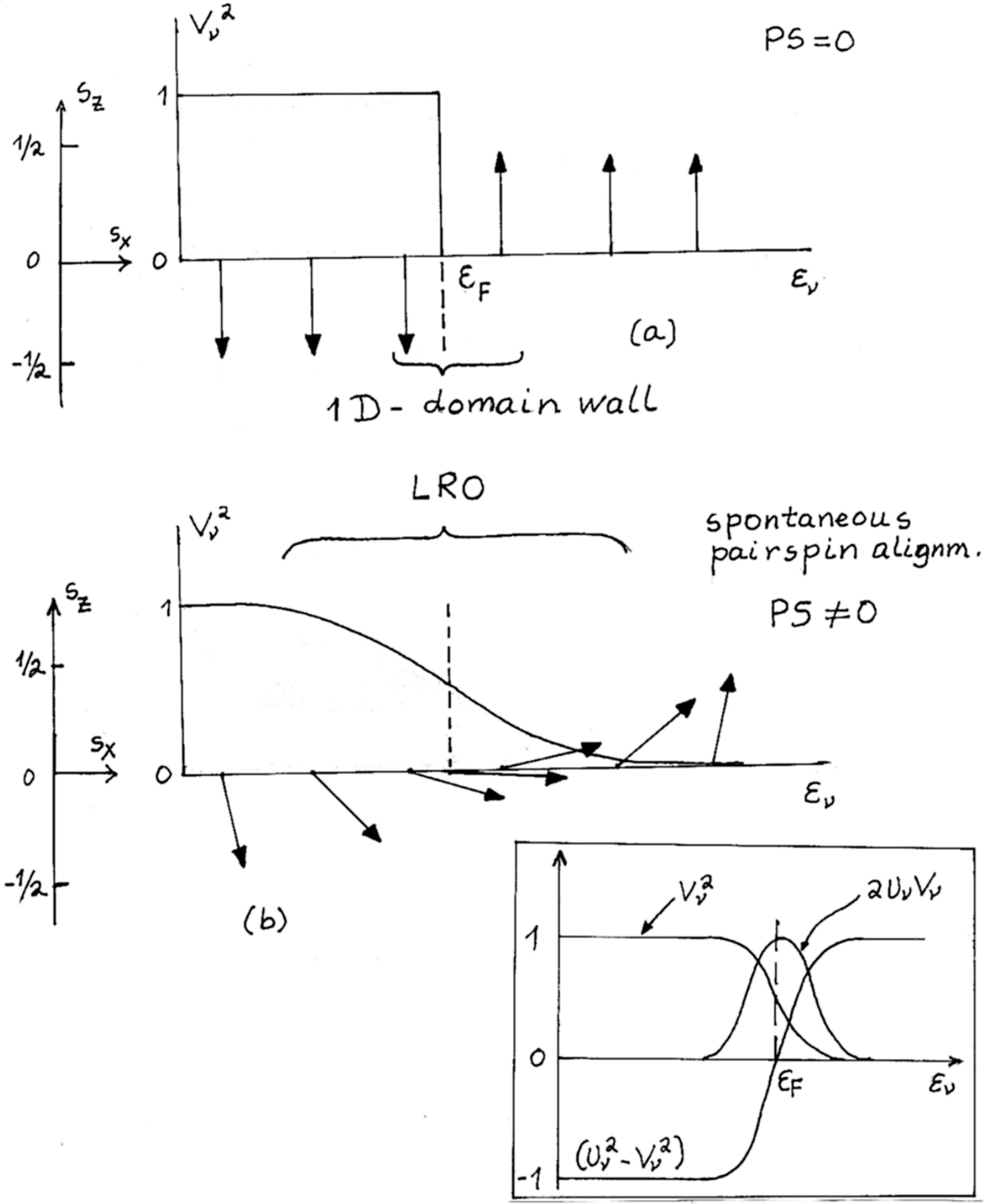}
	\end{center}
	\caption{Nuclear pairspin (PS) or quasispin (QS) analogy to the electron gas \cite{Anderson:58}. The figure illustrates the occupancy $\left( V^2_{\nu}\right)$ and the pairspin magnitude and polarization ($\vec{s}(s_x, s_y, s_z)$, i.e. $s_{z'}(\nu) = \left( U^2_{\nu} - V^2_{\nu} \right) s_z(\nu) + 2 U_{\nu} V_{\nu} s_x(\nu)$), as a function of $\varepsilon_{\nu}$ (see also inset). (a) Non--correlated system ($PS = 0$; $\alpha_0=0$) displaying zero pairspin alignment, can be viewed as a one--dimensional domain wall; (b) the superconducting (nucleon superfluid) ground state displaying Long Range Order (LRO) and a finite value of the total pairspin ($PS\neq 0$; $\alpha'_0 = \sum_{\nu>0} U_{\nu} V_{\nu}$). The pairspin vectors show a gradual rotation in the ($x$,$z$)--polarization plane, like a domain wall (see Fig. 8).}
\end{figure}




\begin{figure}[h]
	\begin{center}
		\includegraphics[width=0.45\textwidth]{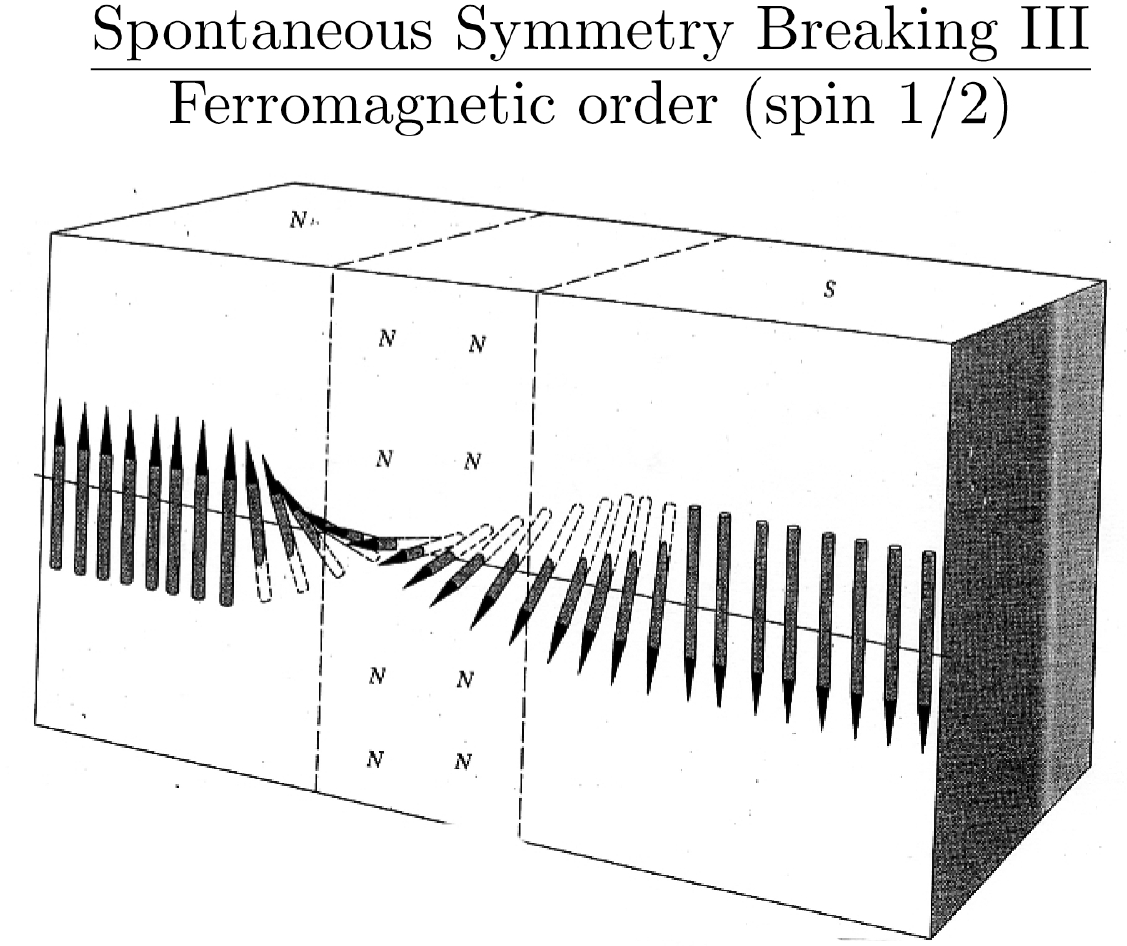}
	\end{center}
	\caption{\protect Below the Curie temperature, ferromagnets acquire a magnetic moment$\rightarrow$spontaneous symmetry breaking of rotational invariance. Emergent property: domain walls. Bloch wall in a crystal is the transition layer that separates adjacent regions (domains) magnetized in different directions, that is, in which North (N) and South (S) ``poles'' are inverted. Reprinted with permission from C. Kittel, Introduction to solid state physics, John Wiley and Sons, 6th ed. (1986) p.453.}
\end{figure}

Getting inspiration from Phil Anderson's quasispin model of superconductivity (\cite{Anderson:58} , see Fig. 7) developed to describe the normal-superconducting phase transition in metals, but also used at profit to make it intuitive the spontaneous breaking of gauge invariance associated with superfluid nuclei in terms of pairspin alignment, and associated privileged orientation in gauge space, one can represent each amino acid in a polymer through a quasispin of magnitude $19/2$ and thus displaying 20 different projections, each representing one of the twenty commonly occurring residues in nature. The lowering of evolutionary temperature in the space of sequence, starting from random sequences (i.e. sequences displaying probability 1/20 of realization of any of the quasispin projections), evolution has aligned a grand total  of 20--25\% of all amino acids \footnote{It is of notice that, was it not because of its very uneven distribution along the linear chain, connected as we now know with the LES mechanism of folding, the number of highly conserved amino acids found in  families of proteins displaying the same native fold ($\lesssim 25\%$ of the total number of amino acids forming the protein) is not different from that which can be aligned between two proteins not related by evolution, i.e. between analogous proteins (cf. \cite{Rost:97,Tiana:00}).} (similar to the way spins are aligned in a ferromagnet below the Curie temperature, see Fig. 8), associated with few (2--3) groups of 10--15 amino acids each (Fig. 9). In other words, quasispin alignment of small groups of residues (similar to magnetic domains) all corresponding to more or less hydrophobic species \cite{Broglia:arxiv}. Such groups give rise, already in the denatured state, to Local Elementary Structures (LES) which direct  folding and, upon docking give rise to the postcritical folding nucleus ((FN)${}_{pc}$), the minimum set of native contacts which inevitable grows into the native state N (Figs. 9 and 10 (A); cf. e.g. \cite{Broglia:01,Broglia:98,Tiana:09} and refs. therein; see also \cite{Broglia:arxiv}).

This microscopic scenario of symmetry breaking in information space, associated with a second order-like phase transition (see \cite{Ramanathan:94} and refs. therein, see also \cite{Banavar:arxiv}), which, arguably, is at the basis of the phenomenon of protein folding and thus of metabolism and consequently, of (one of two possible) origins of life on earth\cite{Dyson:99} , has a very direct consequence. The likely existence of non-conventional folding inhibitors of e.g. viral target proteins, in particular of the HIV--1--PR (an enzyme which plays a central role in the HIV life cycle). These inhibitors are not expected to create resistance\cite{Broglia:08,Broglia:09,Broglia:09brev,Broglia:06brev}), and, arguably, lead to small side effects (see below, in particular Figs. 10 (B) and (C)).

\begin{figure}[htb!]
	\begin{center}
		\includegraphics[width=0.8\textwidth]{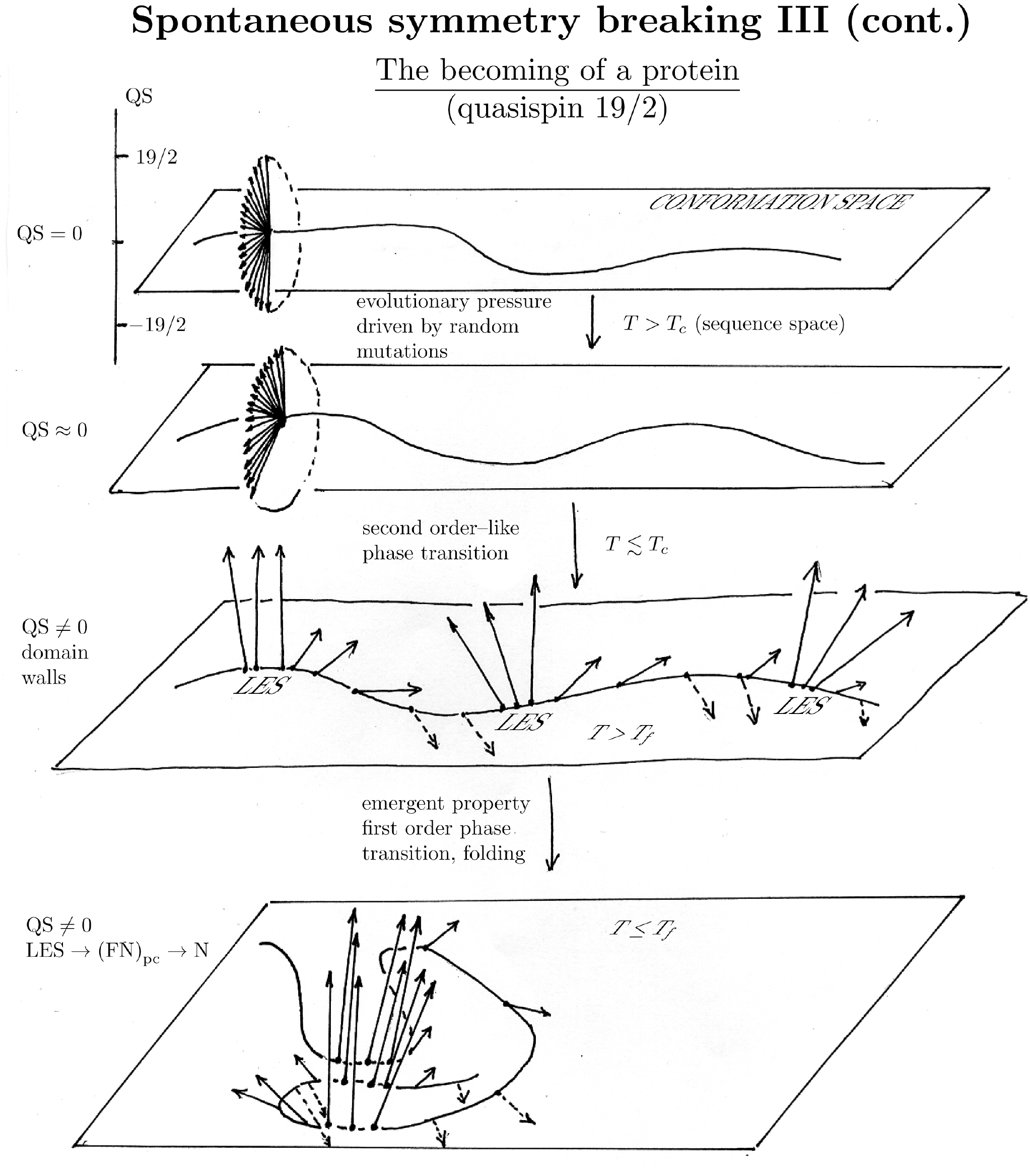}
		\caption{Schematic representation of protein evolution from a random sequence of amino acids into a folded, active enzyme, in terms of the alignment of a $19/2 \;$  Quasi Spin (QS) in sequence (information) space, each of the 20 projections representing one of the twenty amino acids existing in nature. \footnotemark}
	\end{center}
\end{figure}

Within this context, it is of notice that the phenomenon at the basis of the presence of LES in protein folding and stability, of the pigmy resonance in the binding of the halo Cooper pair in $^{11}$Li, of pairing rotational bands in the spectra of atomic nuclei, and of magnetic domains in ferromagnets, is intimately connected with broken symmetry phenomena. Thus the label \textbf{Spontaneous Symmetry Breaking} appearing in the headings of Figs. 2,5 (see also 7) and 8-9.

LES, which can be viewed as incipient, virtual secondary structures, already  present with varied degree of stability in the denaturated state (see Fig. 10 (A)) of the protein \cite{Anfinsen:73}, control not only folding but also aggregation \cite{Broglia:98,Shakhnovich:99}. To make virtual LES become real, one can intervene and block the folding process with peptides displaying identical sequence of LES of the protein under study \cite{Broglia:03}. Such peptides, called p-LES, can bind a complementary LES leading to misfolding and thus competing with productive folding \cite{Anfinsen:73,Broglia:08,Pincus:92}. Circular dichroism is consistent with such a scenario \cite{Broglia:06,Broglia:05prot} , while NMR indicates that the only amino acids which give a signal similar to that associated with the native state of the protein are those which bind in the native state to the LES of which the peptide p-LES is a replica \cite{Caldarini:09}.

In other words p-LES plays the role, in the protein folding process, of the hydrogen atom concerning ZPF of the QED vacuum, and of the excitation of the $\vert 1/2^{-}; 2.69 \textrm{MeV} \rangle$ state concerning phonon mediated pairing in nuclei. Thus the label \textbf{Virtual processes become real}, appearing in the heading of Figs 1, 4 and 10. 

\begin{figure}[htb!]
	\begin{center}
		\includegraphics[width=0.72\textwidth]{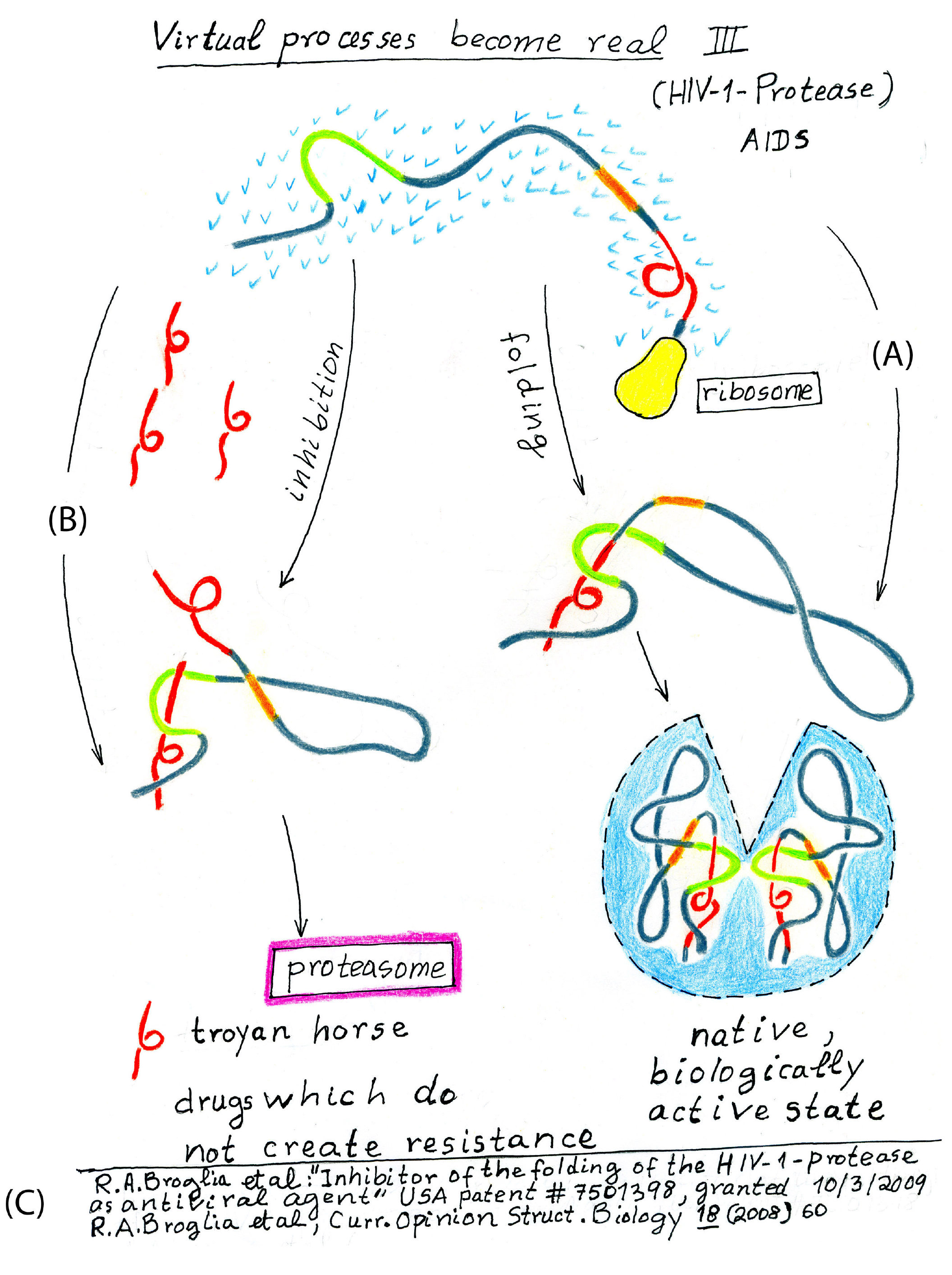}
		\caption{Cartoon representation of the folding (A) and of the folding inhibition (B) of a protein (homodimer) (Cf. \cite{Broglia:arxiv} and \cite{Broglia:08,Broglia:06,Broglia:05prot,Rusconi:07,LoCicero:07,Ferramosca:08,Ferramosca:09} and refs. therein). LES are colored green, orange and red. This last color is also used for the p-LES displaying the same sequence as the red segment of the target protein. Solvent (blue $\bigvee$) is supposed to be all around the protein and the peptide.}
	\end{center}
\end{figure}
\footnotetext{Below the critical evolutionary temperature $T_c$, QS alignment takes place indicating that particular sites of the polypeptide chain become occupied by specific, and thus highly conserved (essentially single projection), as a rule hydrophobic, amino acids. The most prominent emergent property of this (ferromagnetic, Curie-like) phase transition \cite{Broglia:arxiv,Ramanathan:94}, breaking symmetry in information (linear amino acid sequence) space is associated with the formation, already in the denaturated state of a newly expressed protein, of few Local Elementary Structures (LES), virtual secondary like structures flickering in and out of the native conformation (\cite{Broglia:01,Broglia:98,Tiana:09} see also \cite{Anfinsen:73}), which provide molecular recognition to direct folding on short call. In fact, upon docking in the most stable conformation, they build the $(FN)_{pc}$, structure closely related to the transition state but more committed to folding, and thus a domain wall which eventually forces the remaining amino acids to fold in place leading to a biologically active protein. Peptides displaying sequences identical to those of LES can be used to inhibit the folding process, and thus inhibit biological activity (Fig. 10(B)). Due to the fact that in a cell, misfolded proteins are degraded by the proteasome, this reversible event can become permanent. Because to avoid the effect of a p-LES, the pathogen expressing the protein has to introduce mutations in the LES (mutations which obliterate, or in the best case blur, the QS alignment in information space, essentially canceling the workings of evolution and thus forcing the protein to become again a random polymer) folding inhibitors are likely to be leads of drugs which do not create resistance \cite{Broglia:08,Broglia:09,Broglia:06brev,Broglia:09brev,Broglia:06,Broglia:05prot,Rusconi:07,LoCicero:07,Ferramosca:08,Ferramosca:09}. } 

As schematically indicated in this last figure, p-LES can become leads to non-conventional (folding) inhibitors, drugs which likely do not create resistance. In fact, the only way a target protein can avoid a p-LES to bind the complementary LES is by mutating hot amino acids stabilizing LES. But such an event will lead to denaturation (the mutated protein will not able to fold and thus to become biologically active). This does not mean that a target protein cannot develop resistance. It only means that to do so a concerted mutation of a large number of amino acids has to take place in a single step, an event which is very unlikely \cite{Tiana:00,Tiana:09}.

To shed light on the above issues, assuming the target protein to be an enzyme like e.g. the HIV--1--Protease, choice target in the fight against AIDS \cite{Tomaselli:00}, folding inhibition and thus loss of enzymatic activity has to be measured in activity assays (making use of the expressed and purified enzyme \cite{Broglia:06} by itself and in the presence of p-LES), \textit{ex vivo} in acute and in chronically infected cells \cite{Rusconi:07,LoCicero:07}, \textit{in vitro} passages over long periods of time \cite{Ferramosca:08,Ferramosca:09} and in living organisms, a program which is well under way at FoldLESs S.r.l.\cite{Foldless}. Within this context one can mention that there are many ways one (in the present case a theoretical nuclear physicists) can interact with pharmaceutical companies to have such an expensive and ``risky'' project supported, let alone funded. Our experience has been that a very attractive mode is through a University-Pharmaceutical spin off like FoldLESs S.r.l., a joint venture of the University of Milan and of the pharmaceutical multinational Rottapharm-Madaus, a venture that was triggered by nuclear and statistical physicists \cite{Foldless}.

Discussions and collaboration through the years with G. Tiana and E. I. Shakhnovich are gratefully acknowledged. Comments of F. Barranco and E. Vigezzi are much appreciated.


\begin{thebibliography}{99}

\bibitem{Anderson:72}
P.W. Anderson, \emph{More is different}, Science \textbf{177} (1972) 393--396.

\bibitem{Anderson:95}
P.W. Anderson, \emph{Physics, the opening to complexity}, Proc. Natl. Ac. Sci. USA \textbf{92} (1995), 6653--6654.

\bibitem{Bohr:53}
A. Bohr and B.R. Mottelson, \emph{Collective and individual-particle aspects of nuclear structure}, Kgl. Dan. Vidensk. Selsk. Mat. Fys. Medd. \textbf{27}, n. 16 (1953).

\bibitem{Mottelson_Varenna} B.R. Mottelson, \emph{Selected topics in the theory of collective phenomena in nuclei}, Procs. of the International School of Physics "Enrico Fermi",
XV Course, ed. G. Racah, Nuclear spectroscopy, Academic Press (1962), 44--99.

\bibitem{Bohr_Paris} A. Bohr, \emph{Elementary modes of nuclear excitation and their coupling}, Comptes Rendus du Congr\`es International de Physique Nucl\'eaire, Vol I, Editions du CNRS, Paris (1964), 487. 

\bibitem{inhibition} R.A. Broglia and G. Tiana, \emph{Dynamics and conserved regions of the HCV-Protease}, Part I, Stability (static)  and conservation (Dec. 15, 2010);
R.A. Broglia, G. Potel and G. Tiana, \emph{Dynamic, MD simulations} (Dec. 16, 2010); R.A. Broglia and G. Tiana, \emph{ Leads of folding inhibitors of the HCV-PR}, (Nov. 28, 2011),
FoldLESs, S.r.l., internal reports.

\bibitem{Anderson:84}
P.W. Anderson and D.L. Stein, \emph{Broken symmetry, emergent properties, dissipative structures, Llife: are they related?}, in P.W. Anderson, \emph{Basic Notions of Condensed Matter}, Benjamin, Menlo Park, CA (1984) 263--285.

\bibitem{Bes:76}
D.R. B\`es, R.A. Broglia, G.G. Dussel, R.J. Liotta and H.M. Sofia, \emph{The nuclear field treatment of some exactly soluble models}, Nucl. Phys. \textbf{A260} (1976) 1--26.

\bibitem{Bes:76b}
D.R. B\`es, R.A. Broglia, G.G. Dussel, R.J. Liotta and H.M. Sofia, \emph{Application of the nuclear field theory to monopole interactions which include all the vertices of a general force}, Nucl. Phys. \textbf{A260} (1976) 27--76.

\bibitem{Bes:76c}
D.R. B\`es, R.A. Broglia, R.J. Liotta and R.P.J. Perazzo, \emph{On the many--body foundation of the nuclear field theory}, Nucl. Phys. \textbf{A260} (1976) 77--94.

\bibitem{Broglia:76}
R.A. Broglia, B.R. Mottelson, D.R. B\`es, R.J. Liotta and H.M. Sofia, \emph{Treatment of the spurious states in nuclear field theory}, Phys. Lett. \textbf{64B} (1976) 29--32.

\bibitem{Bes:75}
D.R. B\`es and R.A. Broglia, \emph{Equivalence between Feynman--Goldstone and particle--phonon diagrams for finite many--body systems}, in \emph{Problems of Vibrational Nuclei}, Procs. of the Topical Conference on Problems of Vibrational Nuclei, Zagreb, Croatia, Yugoslavia, Eds. G. Alaga, V. Paar and L. \v{S}ips, North Holland, Amsterdam (1975) 1--14.

\bibitem{Bortignon:77}
P.F. Bortignon, R.A. Broglia, D.R. B\`es and R.J. Liotta, \emph{Nuclear field theory}, Phys. Rep. \textbf{30C} (1977) 305--360.

\bibitem{Tomonaga:46}
S.I. Tomonaga, \emph{On a relativistically invariant formulation of the quantum theory of wave fields}, Progr. Theor. Phys. \textbf{1} (1946) 27--42.

\bibitem{Schwinger:48}
J. Schwinger, \emph{On quantum--electrodynamics and the magnetic moment of the electron}, Phys. Rev. \textbf{73} (1948) 416--417.

\bibitem{Schwinger:48b}
J. Schwinger, \emph{Quantum electrodynamics. I: a covariant formulation}, Phys. Rev. \textbf{74} (1948) 1439--1461.

\bibitem{Feynman:49}
R.P. Feynman, \emph{Space--time approach to quantum electrodynamics} Phys. Rev. \textbf{76} (1949) 769--789.

\bibitem{Feynman:49b}
R.P. Feynman, \emph{The theory of positrons}, Phys. Rev. \textbf{76} (1949) 749--759.

\bibitem{Feynman:50}
R.P. Feynman, \emph{Mathematical formulation of the quantum theory of electromagnetic interaction}, Phys. Rev. \textbf{80} (1950) 440--457.

\bibitem{Dyson:49}
F. Dyson, \emph{The radiation theories of Tomonaga, Schwinger and Feynman}, Phys. Rev. \textbf{75} (1949) 486--502.

\bibitem{Bohr:75}
A. Bohr and B.R. Mottelson, \emph{Nuclear Structure, Vol. II}, (Benjamin, New York, 1975).

\bibitem{epj_li11} F. Barranco, P.F. Bortignon, R.A. Broglia, G. Col\`o and E. Vigezzi,
\emph{The halo of the exotic nucleus $^{11}$Li: a single Cooper pair}, EPJ \textbf{A 11} (2001) 385.

\bibitem{Brink:10}
D.M. Brink and R.A. Broglia, \emph{Nuclear superfluidity}, Cambridge University Press, Cambridge, UK (2010) 2nd. ed.

\bibitem{Cooper:56}
L. Cooper, \emph{Bound electron pairs in a degenerate Fermi gas}, Phys. Rev. \textbf{104} (1956) 1189

\bibitem{Bardeen:55}
J. Bardeen and D. Pines, \emph{Electron--phonon interaction in metals}, Phys. Rev. \textbf{99} (1955) 1140

\bibitem{Broglia:05}
R.A. Broglia and A. Winther, \emph{Heavy ion reactions}, Westview Press, Perseus Books, Boulder (2005) 2nd. ed.

\bibitem{Potel:11}
G. Potel et al, \emph{Calculation of the transition from pairing vibrational to pairing rotational regimes between magic nuclei ${}^{100}$Sn and ${}^{132}$Sn 
via two-nucleon transfer reactions}, Phys. Rev. Lett. \textbf{107} (2011) 092501:1--5; Erratum {\bf 108}, (2012) 069904-1

\bibitem{Wimmer:10}
K. Wimmer et al, \emph{Discovery of the shape coexisting $0^+$ state in ${}^{32}$Mg by a two neutron transfer}, Phys. Rev. Lett. \textbf{105} (2010) 252501--4

\bibitem{Josephson:62}
B.D. Josephson, \emph{Possible new effects in superconductive tunneling}, Phys. Lett. \textbf{1} (1962) 251

\bibitem{Bardeen:62}
J. Bardeen, \emph{Tunneling into superconductors}, Phys. Rev. Lett. \textbf{9} (1962) 147

\bibitem{Cohen:62}
M.H. Cohen, L.M. Falicov and J.C. Phillips, \emph{Superconductive tunneling}, Phys. Rev. Lett. \textbf{8} (1962) 316

\bibitem{Anderson:63}
P.W. Anderson and J.M. Rowell, \emph{Probable observation of the Josephson superconducting tunneling effect}, Phys. Rev. Lett. \textbf{10} (1963) 230

\bibitem{Potel:10}
G. Potel et al, \emph{Evidence for phonon mediated pairing interaction in the halo of the nucleus ${}^{11}$Li}, Phys. Rev. Lett. \textbf{105} (2010) 172502:1--4

\bibitem{Tanihata:08}
I. Tanihata et al, \emph{Measurement of the two-halo neutron transfer reaction ${}^{1}$H(${}^{11}$Li,${}^{9}$Li)${}^{3}$H at 3$A$ MeV}, Phys. Rev. Lett. \textbf{100} (2008) 192502

\bibitem{Broglia:73}
R.A. Broglia, O. Hansen and C. Riedel, \emph{Two-neutron transfer reactions and the pairing model}, Advances in Nuclear Physics. \textbf{6} (1973) 287-457. URL: \url{http://merlino.mi.infn.it/repository/BrogliaHansenRiedel.pdf}

\bibitem{Kanungo:prop}
R. Kanungo et al, \emph{Study of nuclear pairing through ${}^{12}$Be(p,t) reaction}, TRIUMF--EEC, Proposal number 1338, approved.

\bibitem{Potel:arxiv2}
G. Potel et al, \emph{Single Cooper pair transfer in stable and in exotic nuclei}, arXiv:0906.4298v3 [nucl--th]

\bibitem{Broglia:arxiv}
R. A. Broglia, \emph{A remarkable emergent property of spontaneous (amino acid content) symmetry breaking}, arXiv:1203.3315v1 [q--bio.BM]

\bibitem{Anfinsen:73}
C.B. Anfinsen, \emph{Principles that govern the folding of protein chains}, Science, \textbf{181}, 223 (1973)

\bibitem{Sanger:52}
F. Sanger, \emph{The arrangement of amino acids in proteins}, Adv. Protein Chem. VII, 1-67, (1952)

\bibitem{Fehrst:99}
A. Fersht, \emph{Structure and mechanism in protein science}, Freeman, New York (1952)

\bibitem{Anderson:58}
P.W. Anderson, \emph{Random--phase approximation in the theory of superconductivity}, Phys. Rev. \textbf{112} (1958) 1900--1916

\bibitem{Rost:97}
B. Rost, \emph{Protein structures sustain evolutionary drift}, Folding Design \textbf{7}, 369 (1958)

\bibitem{Tiana:00}
G. Tiana, R.A. Broglia and E.I. Shakhnovich \emph{Hiking in the energy landscape in sequence space: a bumpy road to good folders}, Proteins \textbf{39}, 244 (2000)

\bibitem{Broglia:01}
R.A. Broglia and G. Tiana, \emph{Hierarchy of events in the folding of model proteins}, J. Chem. Phys. \textbf{114} (2001) 7267--7273

\bibitem{Broglia:98}
R.A. Broglia, G. Tiana, S. Pasquali, H.E. Roman and E. Vigezzi, \emph{Folding and aggregation of designed proteins}, Proc. Natl. Acad. Sci. USA \textbf{95} (1998) 12930--12933

\bibitem{Tiana:09}
G. Tiana and R.A. Broglia, \emph{The molecular evolution of HIV--1 protease simulated at atomic detail}, Proteins \textbf{76} (2009) 895--910  

\bibitem{Ramanathan:94}
S. Ramanathan and E. Shakhnovich, \emph{Statistical mechanics of proteins with ``evolutionary selected'' sequences}, Phys. Rev. \textbf{E 50}, 1303 (1994)

\bibitem{Banavar:arxiv}
J.R. Banavar, T.X. Hoang, F. Seno, A. Trovato and A. Maritan, \emph{Protein sequence and structure: is one more fundamental than the other?}, arXiv: 1204, 2725v1 [q--bio.BM]

\bibitem{Dyson:99}
F. Dyson, \emph{Origins of life}, Cambridge University Press, UK (1999)

\bibitem{Broglia:08}
R.A. Broglia, Y. Levy and G. Tiana, \emph{HIV--1 protease folding and the design of drugs which do not create resistance}, Curr. Opin. Struct. Biol. \textbf{18} (2008) 60--66

\bibitem{Broglia:09}
R.A. Broglia, \emph{Learning to design resistance proof drugs from folding}, Eur. Phys. J. \textbf{D 5}, 137--151, (2009)

\bibitem{Broglia:09brev}
R.A. Broglia, D. Provasi and G. Tiana, \emph{A folding inhibitor of the HIV--1--protease as an antiviral drug}, granted USA patent, \textbf{18} $\sharp 7501398$, (2009)

\bibitem{Broglia:06brev}
R.A. Broglia and G. Tiana, \emph{Method for the identification of protein folding inhibitors}, European Patent, EUR 05015597, PCT/EP2006/0070/8

\bibitem{Shakhnovich:99}
E.I. Shakhnovich, \emph{Folding by association}, Nature Struct. Biol. \textbf{6}, 99, (1999)

\bibitem{Broglia:03}
R.A. Broglia, G.Tiana and R. Berera, \emph{Resistance proof, folding--inhibitor for drugs}, J. Chem. Phys. \textbf{118}, 4754, (2003)

\bibitem{Pincus:92}
M.R. Pincus, \emph{Identification of structural peptide segments in folding proteins}, Biopolymers \textbf{32}, 347, (1992)

\bibitem{Broglia:06}
R.A. Broglia, D. Provasi, F. Vasile, G. Ottolina, R. Longhi and G. Tiana, \emph{Folding inhibitor of the HIV--1 protease}, Proteins \textbf{62}, 928, (2006)

\bibitem{Broglia:05prot}
R.A. Broglia, G.Tiana, L. Sutto, D. Provasi and F. Simona, \emph{Design of HIV--1--PR inhibitors that do not create resistance: blocking the folding of single monomers}, Prot. Sci. \textbf{14}, 2668, (2005)
  
\bibitem{Caldarini:09}
M.Caldarini, F. Vasile, D. Provasi, R. Longhi, G.Tiana and R.A. Broglia, \emph{Identification and chain characterization of folding inhibitors of hen egg lysozyme}, Proteins \textbf{74}, 390, (2009)

\bibitem{Tomaselli:00}
A.G. Tomaselli and R.L. Heinrikson, \emph{Targetting the HIV--protease in AIDS therapy: a current clinical perspective}, Biochem. Biophys. Acta \textbf{1477}, 189, (2000)

\bibitem{Rusconi:07}
S. Rusconi, M. Lo Cicero, S. Ferramosca, F. Sirianni, M. Galli, M. Moroni, A.E. Laface, E. Cesana, A. Clivio, G. Tiana, D. Provasi and R.A. Broglia, \emph{Susceptibility to a non-conventional (folding) protease inhibitor of human immunodeficiency virus Type 1 isolates in vitro}, Proceedings of the International School of Physics "Enrico Fermi", Course CLXV \textit{Protein Folding and Drug Design}, IOS Press, Amsterdam, 293 (2007)

\bibitem{LoCicero:07}
M. Lo Cicero, A.E. Laface, S. Ferramosca, F. Sirianni, E. Cesana, D. Provasi, G. Tiana, M. Galli, M. Moroni, A. Clivio, R. A. Broglia and S. Rusconi , \emph{In vitro activity of a non-conventional (folding) protease inhibitor on human immunodeficiency virus type 1 replication}, Antiviral Therapy \textbf{12}, S19, (2007) 

\bibitem{Ferramosca:08}
S. Ferramosca, M. Lo Cicero, A.E. Laface, F. Sirianni, E. Cesana, D. Provasi, G. Tiana, M. Galli, M. Moroni, A. Clivio, R.A. Broglia and S. Rusconi, \emph{The non-conventional (folding) protease inhibitor blocks the human immunodeficiency virus type-1 replication without evidence of resistance during in vitro passage}, Antiviral Therapy \textbf{13}, A34, (2008)

\bibitem{Ferramosca:09}
S. Ferramosca, M. Lo Cicero, A.E. Laface, F. Sirianni, E. Cesana, D. Provasi, G. Tiana, M. Galli, M. Moroni, A. Clivio, R.A. Broglia and S. Rusconi, \emph{In vitro efficacy of a non-conventional (folding) HIV-1 protease inhibitor without selection of resistance}, Infection, Supplement 2, Abstract \textbf{37}, 23, (2009)

\bibitem{Foldless}
Foldless S.r.l. URL \url{http://merlino.mi.infn.it/news_unimi_foldless_090721.pdf}
\end{thebibliography}
\end{document}